\documentclass{aa}

\usepackage{appendix}

\usepackage[varg]{txfonts}
\usepackage{subfiles}
\usepackage[normalem]{ulem}
\usepackage{xcolor}
\usepackage{placeins}
\graphicspath{{./}{figures/}}

\usepackage[colorlinks=true,citecolor=blue]{hyperref}%
\makeatletter
\renewcommand*\aa@pageof{, page \thepage{} of \pageref*{LastPage}}
\makeatother
\defcitealias{2004ApJ...616..688P}{PK04}
\defcitealias{1995ApJ...451..498M}{Murray et al. 1995}

\defcitealias{2011A&A...536A..49G}{G11}
\defcitealias{2008MNRAS.388..611S}{S08}
\defcitealias{2010MNRAS.404.1369S}{S10a}
\defcitealias{2010MNRAS.408.1396S}{S10b}
\defcitealias{2000ApJ...543..686P}{Proga, Stone \& Kallman 2000}

\newcommand{\mw}{\mathcal{\dot M}_{\rm w}}

\begin{document}
\def\NV{\hbox{{\rm N~}\kern 0.1em{\sc v}}}
\def\PV{\hbox{{\rm P~}\kern 0.1em{\sc v}}}
\def\OVI{\hbox{{\rm O~}\kern 0.1em{\sc vi}}}
\def\OIII{\hbox{[{\rm O~}\kern 0.1em{\sc iii}}]}
\def\CIV{\hbox{{\rm C~}\kern 0.1em{\sc iv}}}
\def\CO{\hbox{{\rm CO}}}
\def\FeII{\hbox{{\rm Fe~}\kern 0.1em{\sc ii}}}
\def\kms{\hbox{km~s$^{-1}$}}

\title{Coordinated X-ray and UV absorption within the accretion disk wind of the active galactic nucleus PG 1126-041}
    \titlerunning{Coordinated X-ray and UV absorption in PG 1126-041}
    \authorrunning{M. Giustini et al.}

\author{M. Giustini
         \inst{1}
         \and
    P. Rodr\'iguez Hidalgo\inst{2,3,4,5}
         \and
    J. N. Reeves\inst{6,7}
        \and
        G. Matzeu\inst{9}
        \and
        V. Braito\inst{7,8,6}
        \and
        M. Eracleous\inst{10,11}
        \and
        G. Chartas\inst{12}
        \and
        N. Schartel\inst{13}
        \and
        C. Vignali\inst{9,14}
        \and
        P. B. Hall\inst{2}
        \and
         T. Waters\inst{15,16}
         \and
         G. Ponti\inst{7,17}
        \and
        D. Proga\inst{15}
        \and
       M. Dadina\inst{14}
        \and
        M. Cappi\inst{14}
        \and\\
        G. Miniutti\inst{1}
        \and
        L. de Vries\inst{18,19}
          }

   \institute{Centro de Astrobiolog\'ia (CAB), CSIC-INTA, Camino Bajo del Castillo s/n, Villanueva de la Ca\~nada, E-28692 Madrid, Spain
   \and
    Department of Physics and Astronomy, York University, 4700 Keele Street, Toronto, ON M3J 1P3, Canada
   \and
    Department of Astronomy and Astrophysics, University of Toronto, 50 St. George Street, Toronto, ON M5S 3H4, Canada
    \and
    Department of Physics and Astronomy, Cal Poly Humboldt, Arcata, CA 95521, USA
    \and
    Physical Sciences Division, School of STEM, University of Washington Bothell, WA, 98011, USA \and
    Department of Physics, Institute for Astrophysics and Computational Sciences, The Catholic University of America, Washington, DC 20064, USA
    \and
INAF, Osservatorio Astronomico di Brera, Via Bianchi 46 I-23807 Merate (LC), Italy
    \and
    Dipartimento di Fisica, Universit\`a di Trento, Via Sommarive 14, Trento 38123, Italy
    \and
    Dipartimento di Fisica e Astronomia ``Augusto Righi'' (DIFA), Universit\`a degli Studi di Bologna, via P. Gobetti 93/2, 40129 Bologna, Italy
    \and
    Department of Astronomy \& Astrophysics, Pennsylvania State University, 525 Davey Lab, University Park, PA 16802, USA
    \and
    Institute for Gravitation and the Cosmos, Pennsylvania State University, University Park, PA 16802, USA
    \and
    Department of Physics and Astronomy, College of Charleston, Charleston, SC, 29424, USA
     \and
     European Space Agency (ESA), European Space Astronomy Centre (ESAC), Camino Bajo del Castillo s/n, Villanueva de la Ca\~nada, E-28692 Madrid, Spain
    \and
    INAF, Osservatorio di Astrofisica e Scienza dello Spazio di Bologna (OAS), via P. Gobetti 93/3, 40129 Bologna, Italy
     \and
    Department of Physics \& Astronomy, University of Nevada, Las Vegas, 4505 S. Maryland Pkwy, Las Vegas, NV, 89154-4002, USA
    \and
    Theoretical Division, Los Alamos National Laboratory, Los Alamos, NM, USA
    \and
     Max-Planck-Institut f\"ur extraterrestrische Physik, Giessenbachstrasse, D-85748, Garching, Germany
    \and
    Amsterdam UMC location University of Amsterdam, Biomedical Engineering and Physics, Meibergdreef 9, Amsterdam, 1105 AZ, The Netherlands
    \and
    Informatics Institute, University of Amsterdam, Amsterdam, The Netherlands
    }

   \date{Received XXX; accepted XXX}


   \abstract
   {Accretion disk winds launched close to supermassive black holes (SMBHs) are a viable mechanism to provide feedback between the SMBH and the host galaxy.
   }
   {We aim to characterize the X-ray properties of the inner accretion disk wind of the nearby active galactic nucleus (AGN) PG 1126-041, and to study its connection with the ultraviolet (UV)-absorbing wind.}
   {We perform spectroscopic analysis of eight \textit{XMM-Newton} observations of PG 1126-041 taken between 2004 and 2015, using both phenomenological models and the most advanced accretion disk wind models available. For half of the dataset, we can compare the X-ray analysis results with the results of quasi-simultaneous, high-resolution spectroscopic UV observations taken with the Cosmic Origins Spectrograph (COS) on board the \textit{Hubble Space Telescope}. }
   {The X-ray spectra of PG 1126-041 are complex and absorbed by ionized material which is highly variable on multiple time scales, sometimes as short as 11 days.
   Accretion disk wind models can account for most of the X-ray spectral complexity of PG 1126-041, with the addition of massive clumps, represented by a partially covering absorber.
   Variations in column density ($N_H \sim 5-20 \times 10^{22}$ cm$^{-2}$) of the partially covering absorber drive the observed X-ray spectral variability of PG 1126-041. The absorption from the X-ray partially covering gas and from the blueshifted {\CIV} troughs  appear to vary in a coordinated way.}
   {
The line of sight toward PG 1126-041 offers a privileged view through a highly dynamic nuclear wind originating on inner accretion disk scales, making the source a very promising candidate for future detailed studies of the physics of accretion disk winds around SMBHs.
}
   {}


   \maketitle
%
\section{Introduction}

Mass outflows are a fundamental physical ingredient of active galactic nuclei (AGN).
While mass accretion onto supermassive black holes (SMBHs, with typical black hole masses $M_{BH}\sim 10^{6-10}\,M_{\odot}$) has been long identified as the main physical mechanism powering AGN \citep[e.g.,][]{1984ARA&A..22..471R}, only recently have AGN been recognized to be able to routinely launch powerful mass outflows, which in turn may profoundly affect the galactic environment \citep[e.g., see][for a recent review]{2021NatAs...5...13L}.
In particular, accretion disk winds -- massive outflows launched on sub-parsec scales  -- may account for typical observational signatures of luminous AGN, such as the broad emission and absorption lines in their optical and ultraviolet (UV) spectra  \citep{1995ApJ...451..498M,2000ApJ...543..686P,2004ApJ...616..688P}.
Different from radio jets, which are highly relativistic, highly collimated and are present in a fraction \citep[about 15-20\%,][]{1989AJ.....98.1195K} of AGN, accretion disk winds are near-relativistic, wide-angle flows of matter whose presence has been inferred to be common in luminous AGN through UV and X-ray spectroscopic studies \citep{1991ApJ...373...23W,2009ApJ...692..758G,2010A&A...521A..57T, 2013MNRAS.430...60G,2021ApJ...920...24C, 2023A&A...670A.182M}.

The innermost regions around the central SMBHs are a rich gaseous environment \citep[see, e.g.,][for a review]{2017NatAs...1..679R}, and X-ray and UV spectroscopic observations bring information about the physical conditions of the matter reprocessing the X-ray and UV radiation. In the case of mass outflows, the column density, ionization state, and velocity shift of the matter absorbing photons from the line of sight can be measured, and therefore the properties of accretion disk winds can be inferred.
In the UV spectra, the most spectacular evidence of winds
are the broad absorption lines (BALs) observed to be blueshifted by $\sim 0.01-0.3\,c$ in the {\CIV} ion  in about 10-15\% of optically selected AGN \citep[e.g.,][]{1981ARA&A..19...41W,1991ApJ...373...23W,2006ApJS..165....1T,2009ApJ...692..758G}.
Such winds might be present in most AGN, if the covering fraction $C_f$ of the outflowing matter is less than 1, i.e., does not cover all the solid angle as seen by the continuum source, and/or there is an evolution of the wind properties across cosmic time, depending on the AGN physical properties \citep[e.g.,][]{2008ApJ...672..102G,2019A&A...630A..94G}.
BALs are usually not observed in local Seyfert galaxies, but in luminous, high-redshift AGN, which are thus called BAL QSOs  \citep[e.g.,][]{2022A&A...668A..87V}. In fact, recent spectroscopic observations of luminous AGN at redshift $z\sim 6$ revealed the presence of BALs in  half of the sample \citep{2022Natur.605..244B}.
The mass outflow rates inferred for high-$z$ BAL QSOs is $\sim 30-400$ $M_{\odot}$ yr$^{-1}$ \citep{2017A&A...601A.143F}.
Properties of the wind, such as the mass outflow rate and the geometrical covering fraction -- therefore the energy deposited by the wind in the environment -- are expected to depend on the AGN spectral energy distribution (SED).
The AGN SED depends on fundamental physical properties such as $M_{BH}$ and the Eddington ratio $\dot{m}=\dot{M}_{BH}/\dot{M}_{Edd}$ \citep[e.g.,][]{1999ApJ...516..672H,2009MNRAS.392.1124V,2012MNRAS.425..907J}, with $\dot{M}_{Edd}$ being the mass accretion rate corresponding to the Eddington luminosity $L_{Edd}=4\pi G m_p M_{BH} c /\sigma_T$, where $m_p$ is the proton mass and $\sigma_T$ is the Thomson cross section. Understanding the physics of AGN is therefore likely intimately connected to understanding the physics of their accretion disk winds.

X-ray observations of BAL QSOs are challenging: first, because of the large cosmological redshift of most of the known sources\footnote{The majority of BAL features are identified from ground observations, thus requiring a cosmological redshift of the source $z \gtrsim 1.7$ in order to identify the typical absorption features in {\CIV} (rest-frame transition at $1549\,$\r{A}).}; then, because of the  observed X-ray weakness of AGN showing BAL features \citep[e.g.,][]{2000ApJ...528..637B,2002ApJ...569..641L}. This X-ray weakness could be intrinsic \citep{2014ApJ...794...70L} or caused by the heavy X-ray reprocessing close to the central engine \citep[e.g.,][]{2001ApJ...546..795G,2002ApJ...567...37G,2022ApJ...936...95W}.
The X-ray properties of BAL QSOs have been therefore usually inferred by means of statistical studies of samples of sources with a very low number of X-ray photons detected \citep[e.g.,][]{2001ApJ...558..109G,2006ApJ...644..709G,2008A&A...491..425G,2009ApJ...690.1006F,2009ApJ...692..758G,2012ApJ...759...42S,2019MNRAS.482.1121S}. A few individual sources have been studied in more detail, thanks to either gravitational lensing \citep[e.g.,][]{2002ApJ...579..169C,2003ApJ...595...85C,2009ApJ...706..644C} or a low cosmological redshift, as in the case of some AGN with BAL-like features identified with space-based UV observations \citep[e.g.,][]{2002ApJ...567...37G,2003AJ....126.1159G,2005A&A...433..455S,2010A&A...512A..75S,2018MNRAS.476..943H}. In these cases, moderate-quality spectroscopy can be performed in order to infer at least the general properties of the absorbing material, such as its column density, covering fraction, ionization state, and, importantly, their variations with time \citep[e.g.,][]{2004ApJ...603..425G,2008A&A...483..137B,2011A&A...536A..49G,2021MNRAS.506..343S}.

After the launch of large effective area X-ray telescopes such as \textit{XMM-Newton}, ultrafast outflows (UFOs) -- X-ray absorbing winds with very large column densities, ionization states, and velocity shifts -- have been observed in a growing number of AGN \citep[e.g.,][]{2002ApJ...579..169C,2003ApJ...595...85C,2003ApJ...593L..65R,2003MNRAS.345..705P,2006AN....327.1012C,2010A&A...521A..57T, 2013MNRAS.430...60G,2021ApJ...920...24C,2023A&A...670A.182M}, and the natural question is what is the relationship between these powerful X-ray absorbing winds and the winds absorbing the UV photons. The answer to this question might have important implications for the theoretical models for accretion disk winds in AGN.

The connection between the X-ray and UV-absorbing gas has implications for the theoretical models of accretion disk winds in AGN, as X-ray photons generally inhibit the formation of UV-absorbing winds.
A large column density of matter able to absorb X-ray photons is needed in  radiation-driven disk wind scenarios, in order to prevent the UV-absorbing wind from becoming overionized (and thus failing). This X-ray absorbing gas has been called `shielding gas' or `hitchhiking gas' \citep{1995ApJ...451..498M}; hydrodynamical simulations have shown that such large columns form naturally in the inner regions of the accretion disk atmosphere, where the gas struggles to escape and forms an inner failed wind \citep{2000ApJ...543..686P,2004ApJ...616..688P}. This inner failed wind in fact acts as a filter of X-ray photons for the outermost UV-absorbing wind, which can then be accelerated farther out \citep[see][for the meaning of successful or failed wind]{2021IAUS..356...82G}.
Recently, attempts have been made at linking the UFOs to the UV BALs, showing promising results \citep[e.g.,][]{2021MNRAS.503.1442M}. However, we caution that these results are based on a specific prescription for the gas opacity which is likely very simplified \citep[e.g., see the discussion in Section 4.2 of][]{2020MNRAS.494.3616N}.
Future hydrodynamical simulations of AGN winds should take into account a realistic treatment of the opacity in the flow, a task that is at the moment beyond the computational possibilities available.

To first approximation, the observed properties of accretion disk winds are inferred using one-dimensional (1D) photoionization codes and simple spherically symmetric geometries, assuming a constant velocity of expansion \citep[e.g.,][]{2012MNRAS.422L...1T,2013MNRAS.430...60G}.
Albeit necessary as a first step of interpretation of the observational results, these 1D spherically symmetric scenarios have implications that are possibly not appropriate for treating realistic accretion disk winds around SMBHs. For example, when multiple absorption troughs of the same ionic species are observed at different velocities, these are usually explained by multiple radial zones of the wind. This might not necessarily be the case if the geometry of the wind is not radial and complex dynamical effects on the wind are taken into account \citep[e.g.,][]{2012ApJ...758...70G}.
The same caveats apply to radial distances estimated using the photoionization approximation and the definition of ionization parameter $\xi=L_{ion}/nR^2$ \citep{1969ApJ...156..943T}, where $L_{ion}$ is the ionizing luminosity, $n$ is the gas density, and $R$ is the radial distance of the absorbing gas parcel from the source of ionizing photons: multiple ionization states of the gas might actually co-exist inside a wind at the same radial distances, as demonstrated by hydrodynamical simulations \citep{2021ApJ...914...62W}.

The physical properties of accretion disk winds around SMBHs were first studied with hydrodynamical simulations by \citet{2000ApJ...543..686P} and \citet{2004ApJ...616..688P}, and these simulations showed similarities between the theoretical absorption line profiles and the observed UV properties of BAL QSOs.
Synthetic X-ray spectra based on the output of the hydrodynamical simulations of \citet{2004ApJ...616..688P} were first presented by \citet{2009ApJ...694....1S} using the one-dimensional photoionization code \texttt{XSTAR} \citep{2001ApJS..133..221K} and later by \citet{2010MNRAS.408.1396S} using Monte Carlo radiative transfer methods.
Despite being based on the output of hydrodynamical simulations that gave predictions for the UV BALs, these studies also showed remarkable agreement between some of the spectral features predicted and the X-ray UFO features observed in luminous AGN.
However, these spectral simulations refer to one single point in the parameter space $[M_{BH},\dot{m}]$
\citep[i.e., the $M_{BH}=10^8\,M_{\odot},\dot{m}=0.5$ input parameters used in the hydro-simulations of][]{2004ApJ...616..688P}, while observations of AGN show that a wide range of parameters are at play.

Radiative transfer calculations based on parametrized (non-hydrodynamical) disk wind models are much faster to compute and have been presented, e.g., in \citet{2001ApJS..133..221K}, \citet[][hereafter \citetalias{2008MNRAS.388..611S}]{2008MNRAS.388..611S}, \citet[][hereafter \citetalias{2010MNRAS.404.1369S}]{2010MNRAS.404.1369S}, and \citet{2018A&A...619A.149L}.
These simulations allow us to probe the parameter space in a physically simplified but computationally much faster way and are needed in order to start constraining the main properties of AGN disk winds, leaving behind the use of simple 1D spherical approximations.
Future dedicated hydrodynamical simulations of accretion disk winds will focus on interesting points in the parameter space found through parametrized wind models studies. These are thus fundamental steps in order to gain a more realistic physical picture of the inner accretion and ejection flows around SMBHs.

An analysis of the X-ray spectra of the X-ray luminous AGN PDS 456, I Zw 1, and MCG-03-58-007, modeled within the disk wind scenario of \citetalias{2008MNRAS.388..611S} and \citetalias{2010MNRAS.404.1369S}, demonstrated a better agreement between the data and the model compared to the use of 1D photoionization codes \citep{2014ApJ...780...45R,2019ApJ...884...80R, 2022ApJ...926..219B}. This was also the case for PG 1448+273, where the blueshifted Fe K shell trough was successfully modeled by \citet{2021A&A...645A.118L} with the disk wind model \texttt{WINE} of \citet{2018A&A...619A.149L}.
In this work, we present the results of a similar experiment on the X-ray weak AGN PG 1126-041, which is known to host BAL-like features in the UV band and both a partially covering absorber and an UFO in the X-ray band.

PG 1126-041 is a low-redshift \citep[$z=0.062$,][]{2009MNRAS.399..683J} active galaxy with an optical magnitude of $M_B=-22.8$ \citep{1983ApJ...269..352S}, very close to the threshold ($M_B=-23$) historically used to divide Seyfert galaxies from the more luminous quasars, or quasi-stellar objects (QSOs).
PG 1126-041 is ``almost'' a Narrow Line Seyfert 1 galaxy (NLS1), having very strong {\FeII} and very weak {\OIII} emission and a full-width-at-half-maximum (FWHM) of the H$\beta$ emission line of 2150 km s$^{-1}$: only slightly larger than the classical FWHM $< 2000$ km s$^{-1}$ used to define NLS1s \citep{1985ApJ...297..166O}.
The black hole mass is estimated to be $M_{BH}=1.2\times 10^8 M_{\odot}$ by \citet{2007ApJ...657..102D} using velocity dispersion measurements in the {\CO} stellar absorption lines.

The UV spectrum of PG 1126-041 shows blueshifted absorption lines in several transitions. An outflow velocity of $\sim - 5000$ km s$^{-1}$ in the {\CIV} and {\NV} species was reported by \citet{1999MNRAS.307..821W} based on \textit{International Ultraviolet Explorer} (\textit{IUE}) data, while an outflow velocity of $\sim - 2000$ km s$^{-1}$ in the {\NV}, {\OVI}, and {\PV} species was reported by \citet{2022ApJ...926...60V} using more recent observations with the Cosmic Origins Spectrograph on board the \textit{Hubble Space Telescope} (\textit{HST}-COS).
These absorption lines have ionization and velocity similar to BALs but a smaller width ($< 2000$ km s$^{-1}$) and are thus called mini-BALs \citep[e.g., section 7.4 of][and references therein]{2022ApJ...926...60V}.
In the X-ray band, PG 1126-041 has a relatively large flux, being the second X-ray brightest PG QSO with known BAL signatures, with a $0.2-2$ keV observed flux of $\sim 10^{-12}$ erg cm$^{-2}$ s$^{-1}$. There are also clear spectral signatures of strong reprocessing of the nuclear X-ray emission by multiple highly variable ionized gas phases, including a mildly-ionized partially covering absorber and a high-velocity ($\sim - 16,500$ km s$^{-1}$), highly ionized component detected with XMM-Newton \citep[][hereafter \citetalias{2011A&A...536A..49G}]{2011A&A...536A..49G}.
A wind is also detected in the H$\alpha$ and {\OIII} optical emission lines, observed with the \textit{Very Large Telescope} (\textit{VLT}) Multi Unit Spectroscopic Explorer (MUSE) adaptive optics to have a blueshift of a few tens to a few hundreds km s$^{-1}$ on kpc-scales \citep{2020A&A...644A..15M}.
The line of sight toward the active nucleus of PG 1126-041 is thus privileged: it offers a view of the nuclear wind on multiple radial scales, in particular of the accretion disk wind originating on UV-emitting and UV-absorbing scales, and also of the disk wind originating on X-ray emitting  and absorbing scales.
It might therefore hold important clues on whether such winds are disconnected or to what extent they are connected.

In this work, we present an extension of the work of \citetalias{2011A&A...536A..49G}, in which we double the number of X-ray observations of PG 1126-041, extending the timeline of the study to 11 years (Section \ref{sec:OBS}); test the latest spectral models of accretion disk winds around SMBHs (Sect. \ref{sec:XRAY}); and present high-resolution spectroscopic observations of the {\CIV} line profile partially overlapping with the X-ray observations (Sect. \ref{sec:XUV}). We discuss our results in Sect. \ref{sec:DISCU}, and present the conclusions in Sect. \ref{sec:CONCLU}.
A cosmology with $H_0 = 70$ km s$^{-1}$ Mpc$^{-1}$, $q_0=0$, and $\Omega_{\Lambda}=0.73$ is adopted throughout the paper \citep{2016A&A...594A..13P}.
The corresponding luminosity distance to PG 1126-041 is 278.4 Mpc. Errors and error bars are at the $1\sigma$ level, unless otherwise stated.

\section{Observations and data reduction\label{sec:OBS}}

\begin{table*}[ht!]
\caption{Log of the \textit{XMM-Newton} observations of PG 1126-041  \label{T_OBS}}
\centering
\begin{tabular}{ccccccc}
\hline\hline
Name & OBSID &  Start date           & End date & t$_{exp}$ & t$_{net}^{pn}$ & ct s$^{-1}$  \\
(1) & (2) & (3) & (4) & (5) & (6) & (7) \\
\hline
 2004 & 0202060201 & 2004-12-31 01:12:42  & 2004-12-31 10:36:22 & 33.8 &  29.1 & $0.187\pm{0.003}$ \\
 2008A & 0556230701  & 2008-06-15 06:53:14 & 2008-06-15 15:36:00  & 31.4 & 4.0 & $0.282\pm{0.009}$ \\
 2008B & 0556231201 & 2008-12-13 20:05:14  & 2008-12-13 23:23:47  & 11.9 & 3.6 & $0.53\pm{0.01}$ \\
 2009 & 0606150101 & 2009-06-21 06:57:04 & 2009-06-22 19:45:27 & 132.4 & 71.1 & $0.188\pm{0.002}$ \\
 2014A & 0728180201 & 2014-06-01 06:07:18 & 2014-06-01 16:05:38 & 35.9 & 16.3 & $0.093\pm{0.003}$ \\
 2014B & 0728180301 & 2014-06-12 20:43:22 & 2014-06-13 03:06:42 & 23.0 & 17.3 & $0.211\pm{0.004}$ \\
 2014C & 0728180401 & 2014-06-28 18:34:32 & 2014-06-29 02:21:12 & 28.0 & 21.4 & $0.154\pm{0.003}$ \\
 2015 & 0728180501 & 2015-06-14 07:04:49 & 2015-06-14 12:04:49 & 18.0 & 12.9 & $0.251\pm{0.004}$ \\
\hline
\end{tabular}
\tablefoot{(1) Name used in the article; (2) Observation ID; (3) Starting date of observation (yyyy-mm-dd hh:mm:ss UTC); (4) Ending date of observation (yyyy-mm-dd hh:mm:ss UTC); (5) Exposure time (ks); (6) Net EPIC-pn exposure time after flaring background removal (ks); (7) EPIC-pn net count rate in the 0.3-10 keV band. }
\end{table*}

We analyzed eight \textit{XMM-Newton} pointed observations of PG 1126-041 performed between 2004 and 2015, with exposure times ranging between 10 and 133 ks (Principal Investigator (PI): M. Giustini for all the observations except for the first one, with PI N. Schartel). Of these observations, the first half has been already published by \citetalias{2011A&A...536A..49G}.
The second half of the dataset was taken quasi-simultaneously with observations performed with the \textit{HST}-COS.
Details  about  the  COS  observations  of  PG  1126-041  will  be reported  in  a  companion  article  (Rodr\'iguez  Hidalgo  et  al.,  in prep.).

All the \textit{XMM-Newton} observations  of PG 1126-041 were taken in full frame mode with the medium optical filter except for the 2004 observation, which was performed with the thin optical filter.
The observation data files were processed with the science analysis system (SAS) v.18.0.0 using calibration files generated in January 2021.
Given the low X-ray count-rate of PG 1126-041, pile-up effects are negligible.
The exposure time is also too short for a detection with the high-resolution reflection grating spectrometers (resolving power $R\sim 100-500$), therefore we restricted the X-ray analysis to the European Photon Imaging Camera (EPIC) data ($R\sim 10-50$, see the \textit{XMM-Newton} Users Handbook for details\footnote{]\url{https://xmm-tools.cosmos.esa.int/external/xmm_user_support/documentation/uhb/}}).
These were reprocessed using standard SAS analysis threads, using the tasks \texttt{epproc} and \texttt{emproc} to concatenate the raw EPIC-pn (pn hereafter) and EPIC-MOS (MOS hereafter) events.
A range of thresholds in count rate of $0.5-1.0$ ($0.35-0.5$) ct s$^{-1}$ for high-energy single events (10 keV $< E <$ 12 keV, PATTERN=0) applied to the light curve of the whole field of view was used to filter out background flares from the pn (MOS) data.
Single and double pattern events (PATTERN$\leq$ 4) or up to quadruple pattern events (PATTERN$\leq$12) with quality flags \#XMMEA\_EP and \#XMMEA\_EM, for the pn and the MOS data, respectively, were retained to extract the source and background spectra.
Source+background spectra were extracted from circular regions centered on the source coordinates, with optimal radii determined with the SAS task \texttt{eregionanalyse}  to be between 27'' and 45'', depending on the signal-to-noise ratio (S/N).
Background spectra were extracted from source-free regions of the detector, with circular(annular) shapes in the case of pn(MOS) data.
The background extraction regions were always larger than the source ones; the areas were normalized using the \texttt{backscale} SAS task.
Ancillary response file and response matrix were generated with the \texttt{arfgen} and \texttt{rmfgen} SAS tasks at the source positions. The MOS spectra and response files were combined with the \texttt{epicspeccombine} SAS task.

The \textit{XMM-Newton} observation identifiers (OBSID), date of observation, exposure time before and after the flaring background filtering, and pn net count rate are reported in Table~\ref{T_OBS}.

\section{X-ray spectral analysis\label{sec:XRAY}}

The software \texttt{Xspec}\,\texttt{v.12.12.1} \citep{1996ASPC..101...17A} with the python interface \texttt{pyXspec} \citep{2021ascl.soft01014G} was used for the X-ray spectral analysis, and the $\chi^2$ statistic was employed to assess the goodness of fit and estimate measurement errors.
The pn and MOS spectra were grouped to a minimum number of 30 counts per energy bin  using the task \texttt{ftgrouppha}\footnote{\url{https://heasarc.gsfc.nasa.gov/ftools/}}, to guarantee a minimum number of 20 counts per bin after background subtraction.
The energy range considered for the spectral analysis was $0.3-10$ keV, where the instruments are well-calibrated and their effective area is non-negligible. Bad channels were ignored.

      \begin{figure}[hbt!]
  \centering
  \includegraphics[width=9cm]{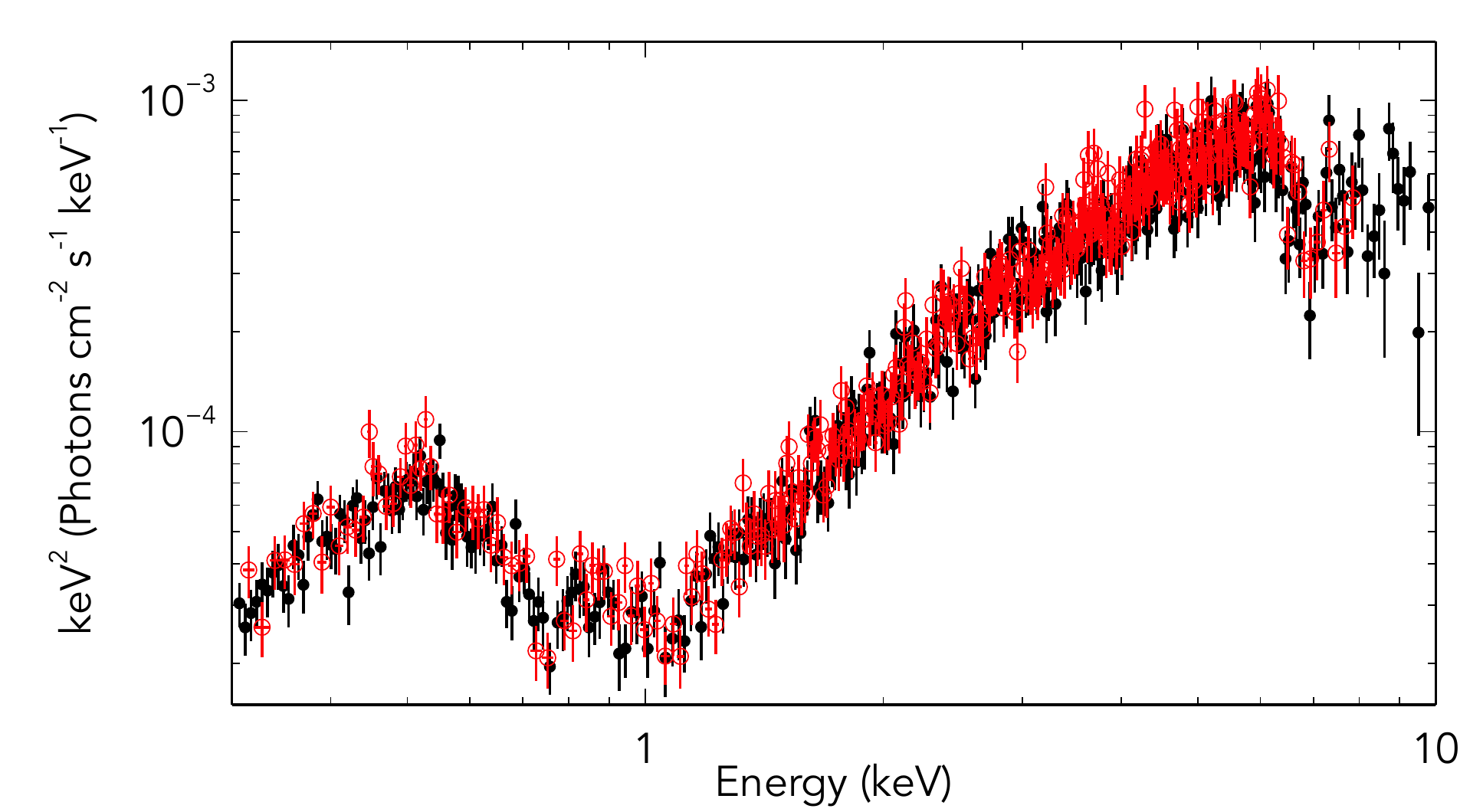}
   \caption{EPIC spectra of PG 1126-041 observed in June 2009. Black circles are pn data, while red open circles are MOS data. The spectra are plotted unfolded against a power law model with $\Gamma=2$.}
   \label{FIG3}%
    \end{figure}

        \begin{figure*}[hbt!]
  \centering
  \includegraphics[height=12.8cm]{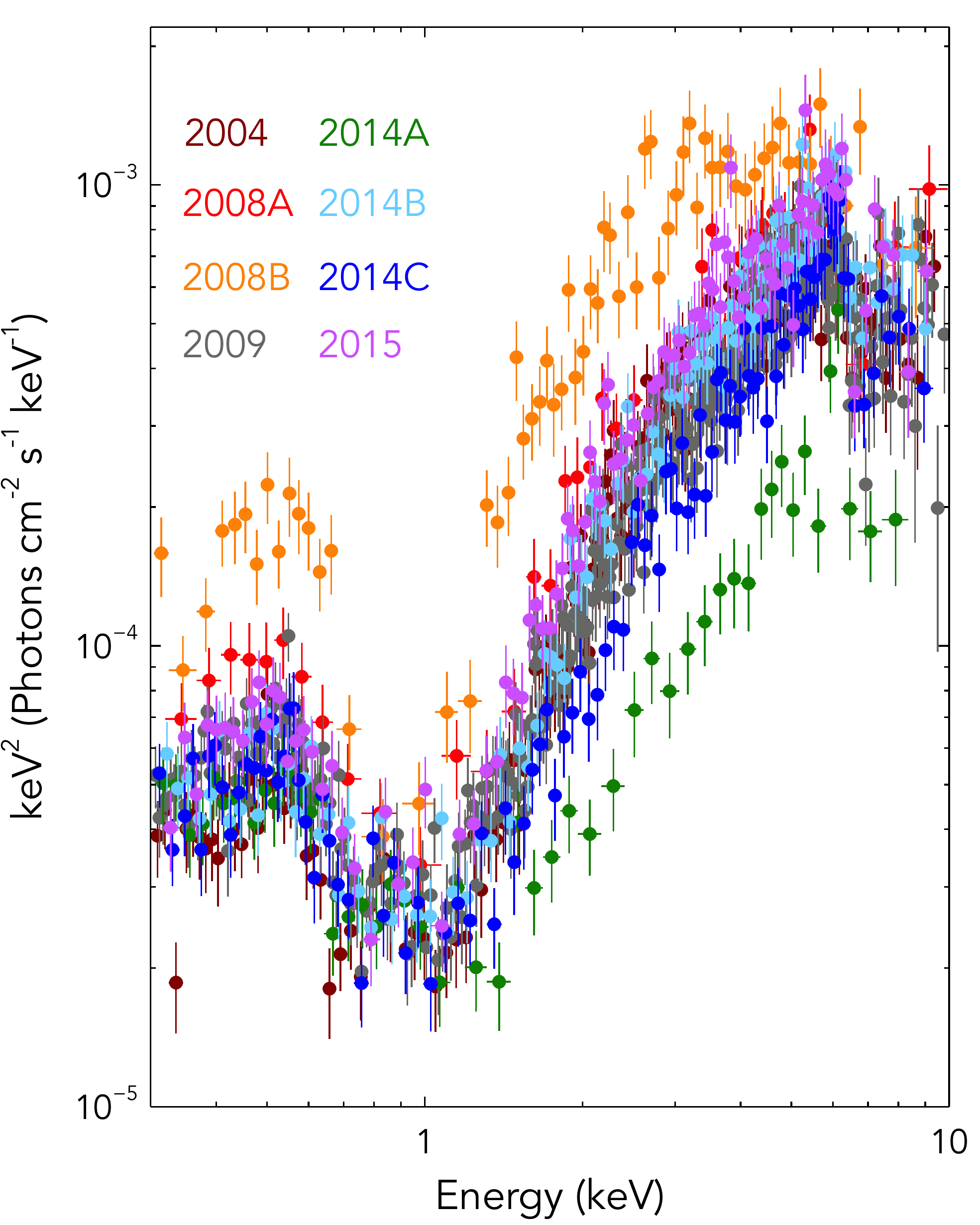}
  \includegraphics[height=12.8cm]{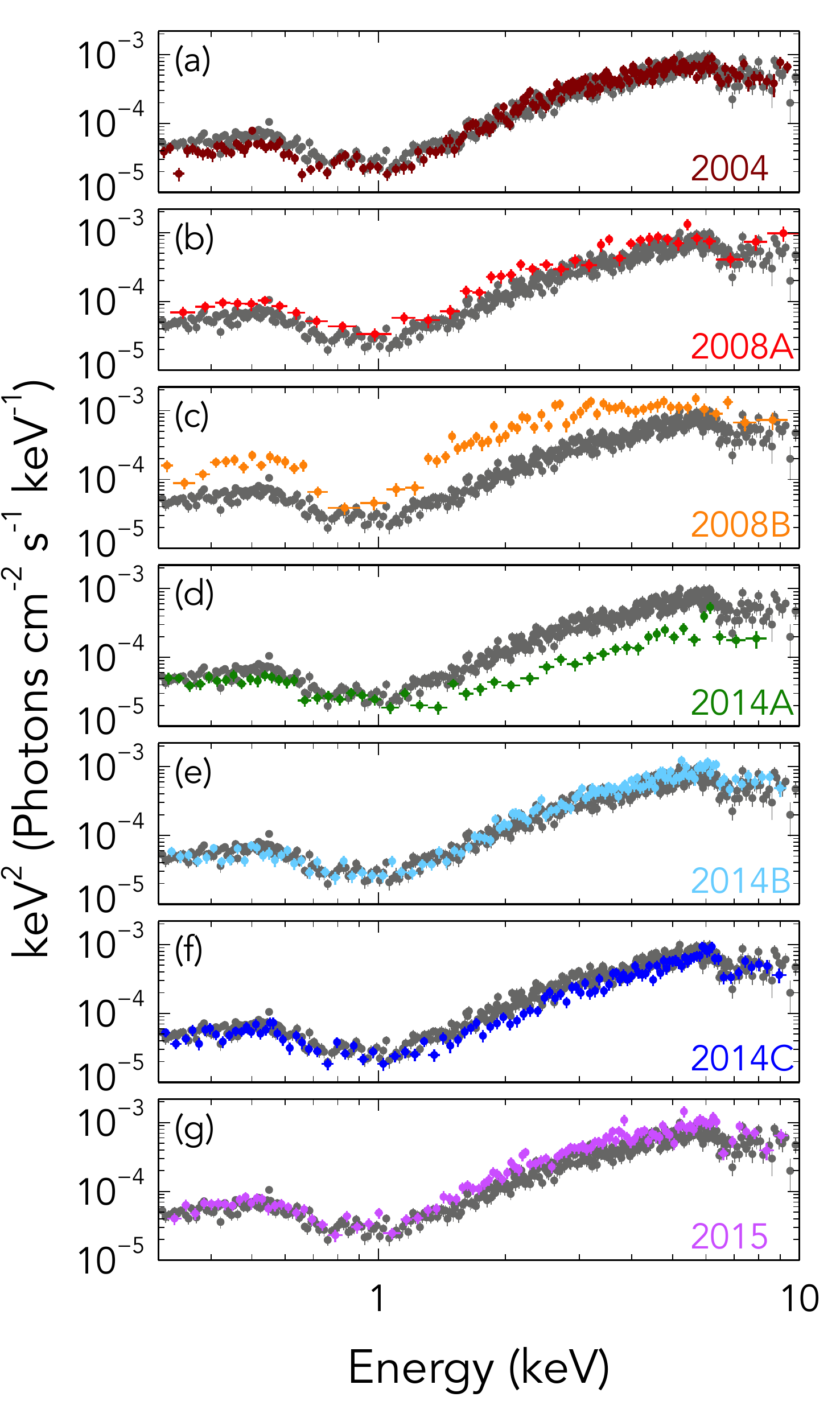}
  \caption{\textit{XMM-Newton} EPIC-pn spectra of PG 1126-041 extracted in the $0.3-10$ keV band in eight different epochs of observation. Left panel:  spectra unfolded against a power law with $\Gamma=2$.
  Right panel: the spectra of each epoch unfolded against a $\Gamma=2$ power law model are plotted individually over the 2009 spectrum as a reference.}
   \label{FIG1}%
    \end{figure*}

The high-S/N EPIC spectra of the 2009 observation of PG 1126-041 are shown in Fig.~\ref{FIG3} unfolded against a power law model with a photon index $\Gamma=2$. The black crosses refer to the pn data, while the red open circles to the merged data of the two MOS cameras. There is good agreement between the data recorded by the two cameras, and in most of the following plots we will only show the pn data. For each epoch of observation, the pn and MOS spectra were fitted to the same model, and uncertainties in cross-calibration between the two instruments were taken into account by adding a multiplicative constant component $C_{\textrm{MOS}}\in [0.8,1.2]$ between the two spectra of each epoch of observation.

A Galactic column density of $N_H^{Gal}=4.35\times 10^{20}$ cm$^{-2}$ along the line of sight \citep{2005A&A...440..775K} was applied to all the spectral models using the \texttt{tbabs} model \citep{2000ApJ...542..914W}.

The $0.3-10$ keV pn spectra of PG 1126-041  observed by \textit{XMM-Newton} between 2004 and 2015 are shown in Figure~\ref{FIG1}, unfolded against a power law model (\texttt{pow} in \texttt{Xspec}) with $\Gamma=2$.
The X-ray spectra of PG 1126-041 are complex, significantly deviating from the simple power law model at all the energies probed by the EPIC cameras; this is the signature of strong reprocessing of the intrinsic continuum emission of PG 1126-041 by material along the line of sight.

 \begin{figure}[ht!]
  \centering
  \includegraphics[width=9cm]{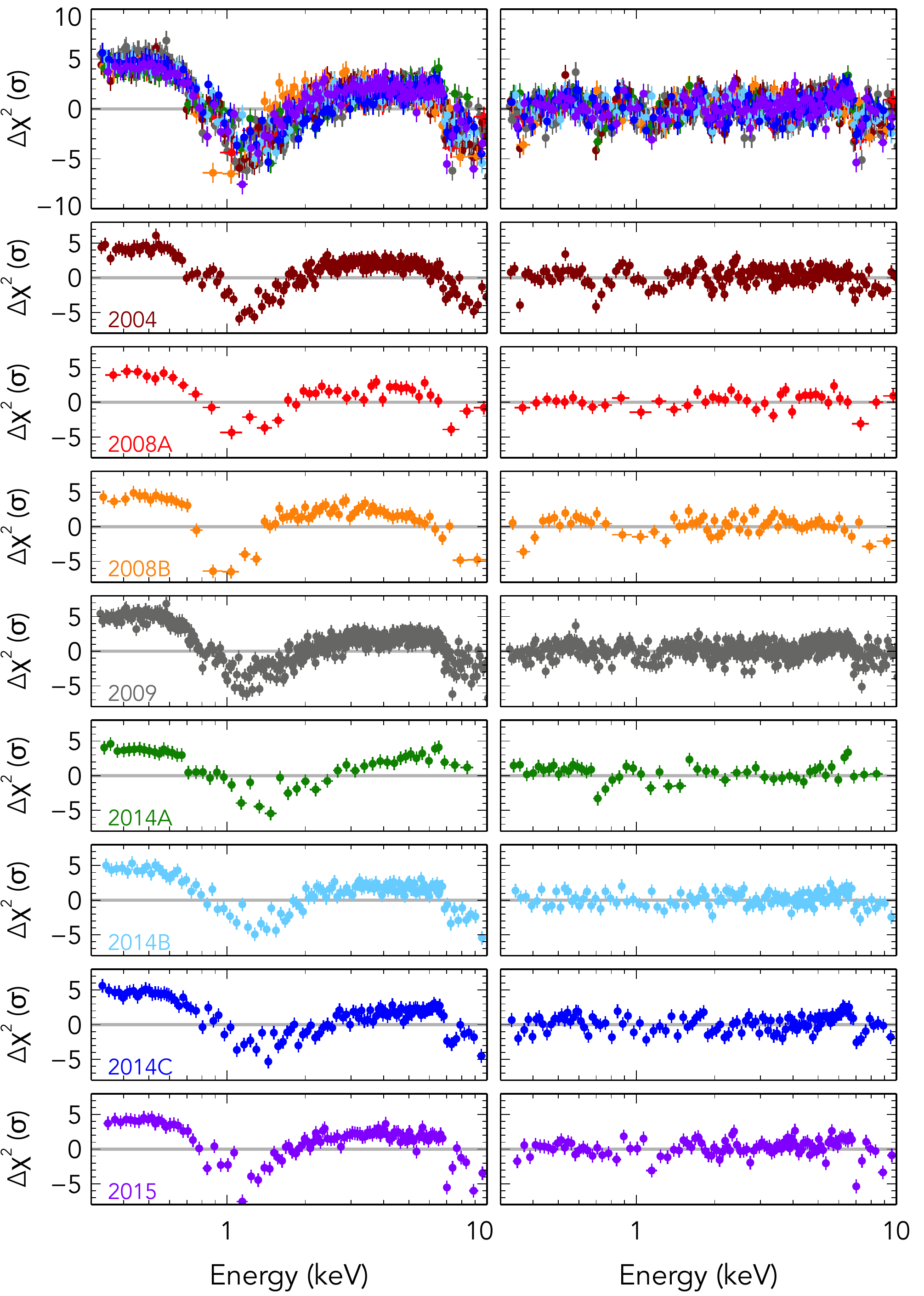}
\caption{EPIC-pn spectral residuals of PG 1126-041 to the power law model \texttt{[pow]} (left column) and to the baseline model \texttt{[(partcov*xstar500)*(pow)]} (right column).
The top panel reports the residuals of all eight epochs together, while the smaller individual panels correspond to different epochs of observation, as listed in Table~\ref{T_OBS}.}
   \label{FIG_RESBASELINE}
    \end{figure}

A fit to a phenomenological power law gives a very flat photon index $\langle\Gamma\rangle \sim 0.7$, compared to both the expected theoretical value \citep[$1.5 <\Gamma < 2.5$,][]{1994ApJ...432L..95H} and the typical value observed in AGN \citep[$\langle \Gamma \rangle \sim 1.8-2$, see e.g.,][]{2005A&A...432...15P}. The  spectral residuals of the eight epochs of observation for the power law model are shown in  the left column of Fig. \ref{FIG_RESBASELINE}.

The X-ray spectral complexity of PG 1126-041 can be reproduced to first order by the addition along the line of sight of a layer of ionized gas that is only partially covering the source of X-ray continuum emission \citepalias{2011A&A...536A..49G}.
We modeled this gas with the code \texttt{XSTAR}, that computes the physical conditions of a geometrically and optically thin shell of gas illuminated by a point source continuum in the $0.1-20$ keV energy range, assuming photoionization equilibrium \citep{2001ApJS..133..221K}.
We used \texttt{XSTAR v2.54a} to generate a grid of spectra assuming a power law ionizing continuum with $\Gamma = 2$ and a luminosity $10^{44}$ erg s$^{-1}$, ionizing a gas shell with a density $n_e = 10^{12}$ cm$^{-3}$ and an intrinsic turbulent velocity\footnote{The value of the turbulent velocity of the absorbing gas was chosen on the basis of the best fit value found using the \texttt{warmabs} version of \texttt{XSTAR} to model the high S/N 2009 spectrum. While \texttt{warmabs} has the advantage of avoiding the intrinsic coarseness of the \texttt{XSTAR} table, it is computationally very expensive and prohibitive to use with our large dataset and our limited computing facilities. Therefore, for the spectral fit and the error computation, we generated \texttt{XSTAR} grids as finely spaced as possible.} $\upsilon_{turb} = 500$ km s$^{-1}$.
The gas column density and ionization parameter were logarithmically sampled at 15 points between $N_H=3\times 10^{22}$ cm$^{-2}$ and $N_H=4\times 10^{23}$ cm$^{-2}$, and at 5 points between $\log\xi=1.5$ and $\log\xi=3$, respectively. From the resulting grid we generated with the task \texttt{xstar2xspec} a multiplicative table to be read into \texttt{Xspec}, which we named \texttt{xstar500}.
In order to account for the absorber only partially covering the source, the
\texttt{xstar500} table was convolved with the \texttt{partcov} model, parametrized by the covering fraction $C_f$.
This is the fraction of the X-ray emission source covered by the absorber, leaving the remaining $(1-C_f)$ X-ray flux unabsorbed.

The model \texttt{(partcov*xstar500)*pow}, hereafter the baseline model, gives a fit statistic $\chi^2/\nu=2779/1923$, where $\nu$ is the number of degrees of freedom, for a joint fit with all the parameters free to vary among the eight epochs of observation.
The average photon index is $\langle\Gamma\rangle=2.04$.
The column density along the line of sight ranges from a minimum $N_H=6.6^{+0.3}_{-0.1}\times 10^{22}$ cm$^{-2}$ during 2008B to a maximum $N_H= 21.0^{+0.4}_{-0.6}\times 10^{22}$ cm$^{-2}$ during 2014A, with an average $\langle N_H\rangle = 1.5\times 10^{23}$ cm$^{-2}$.
The average ionization parameter is $\langle\log\xi\rangle=1.94$, and the covering fraction $\langle C_f\rangle=96.3$.
Fit results are reported in Table~\ref{table:fitBASELINE}, and spectral residuals are shown in the right column of Fig. \ref{FIG_RESBASELINE}.
Negative residuals at $E >6$ keV are visible in all the spectra except for 2014A, while positive residuals between 4 and 6 keV are visible in a half of the spectra.

\begin{table*}
\footnotesize{
\caption{\label{table:fitBASELINE}Spectral fit results for the baseline model \texttt{[(partcov*xstar500)*pow]}.}}
\centering
\begin{tabular}{lllllllll}
\hline\hline
 & 2004 & 2008A & 2008B & 2009 & 2014A & 2014B & 2014C & 2015 \\
 \hline
& \multicolumn{8}{c}{\textit{Power law emission:} \texttt{pow}}\\
$\Gamma$  & $2.05^{+0.02}_{-0.01}$ & $1.92^{+0.03}_{-0.03}$ & $2.03^{+0.06}_{-0.01}$ & $1.95^{+0.01}_{-0.01}$ & $2.35^{+0.04}_{-0.03}$ & $1.97^{+0.02}_{-0.02}$ & $2.05^{+0.02}_{-0.02}$ & $1.97^{+0.02}_{-0.02}$\\
$N_{1\,keV}/10^{-4}$  & $7.6^{+0.1}_{-0.2}$ & $7.3^{+0.3}_{-0.2} $ & $12.7^{+0.2}_{-0.6} $ & $6.0^{+0.1}_{-0.1} $ & $4.4^{+0.2}_{-0.1} $ & $8.0^{+0.2}_{-0.2} $ & $6.7^{+0.1}_{-0.2} $ & $8.2^{+0.3}_{-0.2} $\\
  \multicolumn{9}{c}{ }\\
&\multicolumn{8}{c}{\textit{Partially covering absorber:} \texttt{partcov*xstar500}}\\
$N_H/10^{22}$   & $14.7^{+0.4}_{-0.2}$ & $11.7^{+0.4}_{-0.4}$ & $6.6^{+0.3}_{-0.1}$ & $15.8^{+0.1}_{-0.2}$ & $21.0^{+0.4}_{-0.6}$ & $15.8^{+0.3}_{-0.3}$ & $20.4^{+0.3}_{-0.4}$ & $13.9^{+0.3}_{-0.3}$ \\
$\log\xi$   & $1.91^{+0.01}_{-0.01}$ & $1.97^{+0.01}_{-0.01}$ & $1.74^{+0.02}_{-0.02}$ & $1.98^{+0.01}_{-0.01}$ & $2.04^{+0.03}_{-0.02}$ & $1.94^{+0.01}_{-0.01}$ & $2.02^{+0.02}_{-0.02}$ & $1.94^{+0.01}_{-0.01}$ \\
$C_f$ (\%) & $97.0^{+0.1}_{-0.3}$ & $97.4^{+0.6}_{-0.6}$ & $97.4^{+0.5}_{-0.5}$ & $95.5^{+0.1}_{-0.2}$ & $94.4^{+0.3}_{-0.4}$ & $96.7^{+0.3}_{-0.2}$ & $95.9^{+0.2}_{-0.3}$ & $96.5^{+0.3}_{-0.3}$\\
  \multicolumn{9}{c}{ }\\
&\multicolumn{8}{c}{\textit{MOS cross-calibration constant}} \\
$C$ & $1.13^{+0.04}_{-0.02}$ & $1.08^{+0.05}_{-0.05}$ & $1.18^{+0.09}_{-0.02}$ & $1.07^{+0.02}_{-0.02}$ & $1.07^{+0.06}_{-0.04}$ & $1.06^{+0.04}_{-0.04}$ & $1.00^{+0.05}_{-0.03}$ & $1.10^{+0.04}_{-0.04}$\\
\multicolumn{9}{c}{ }\\
\hline
\textit{Fit statistics} &  & \multicolumn{6}{c}{$\chi^2/\nu=2779/1923$} \\
 &   & \multicolumn{6}{c}{$P_{null}= 2\times 10^{-34}$} \\
\hline
$f_{0.3-10}^{observed}$ & $1.28^{+0.02}_{-0.02}$ & $1.72^{+0.06}_{-0.06}$ & $2.98^{+0.09}_{-0.06 }$ & $1.22^{+0.01}_{-0.01}$ & $0.46^{+0.03}_{-0.01}$ & $1.53^{+0.02}_{-0.03}$ & $1.04^{+0.01}_{-0.04}$ & $1.65^{+0.02}_{-0.05}$\\
$f_{0.3-10}^{unabsorbed}$ & $3.37^{+0.05}_{-0.05}$ & $4.63^{+0.14}_{-0.17}$ & $6.74^{+0.20}_{-0.14}$ & $3.52^{+0.04}_{-0.03}$ & $2.56^{+0.18}_{-0.03}$ & $4.05^{+0.06}_{-0.08}$  & $3.26^{+0.04}_{-0.10}$ & $4.18^{+0.06}_{-0.09}$\\
$L_{2-10}^{unabsorbed}$ & $1.56^{+0.03}_{-0.02}$ & $1.98^{+0.06}_{-0.07}$ & $2.93^{+0.08}_{-0.06}$ & $1.51^{+0.01}_{-0.02}$ & $0.72^{+0.05}_{-0.01}$ & $1.91^{+0.04}_{-0.03}$  &  $1.39^{+0.02}_{-0.04}$ & $1.92^{+0.02}_{-0.05}$\\
\hline
\end{tabular}
\tablefoot{Units: power law normalisation $N_{1\,keV}$ [$10^{-4}$ photons keV$^{-1}$ cm$^{-2}$ s$^{-1}$]; column density $N_H$ [$10^{22}$ cm$^{-2}$]; ionization parameter $\xi$ [erg cm s$^{-1}$]; flux $f$ [$10^{-12}$ erg s$^{-1}$ cm$^{-2}$];  luminosity $L$ [$10^{43}$ erg s$^{-1}$]. The unabsorbed flux is corrected for both intrinsic and Galactic absorption, as is the luminosity. }
\end{table*}

\begin{figure*}
  \centering
  \includegraphics[width=15cm]{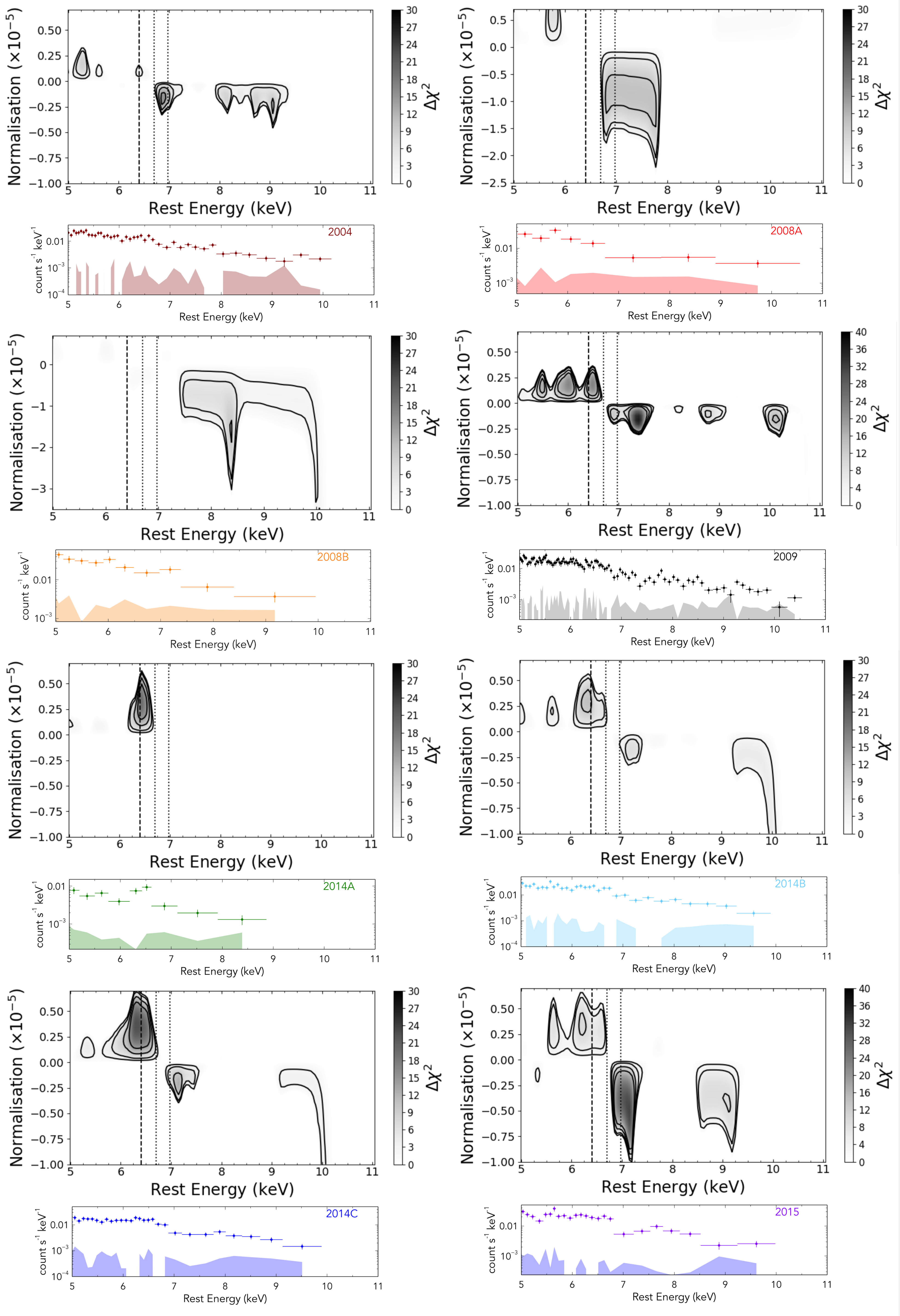}
   \caption{Results of the scan of the $\chi^2$ statistical spaces between $5$ and $11$ keV with a Gaussian line. Top panels: confidence contours at the (from the outermost to the innermost) 68\%, 90\%, 99\%, and 99.9\% significance level  for the centroid energy and normalization of a Gaussian emission/absorption line applied to the baseline model \texttt{[(partcov*xstar500)*pow]}.
    The contours are filled with a color intensity proportional to the $\Delta\chi^2$ represented in the color map at the right of each panel. The three dashed vertical lines mark the rest-frame energy of Fe~I, Fe~XXV, and Fe~XXVI K$\alpha$ transitions. Note the different y-axes in the 2008A and 2008B panels.
   Bottom panels: the observed spectra (filled circles) and the corresponding background (shaded area).}
   \label{FigLINES}%
    \end{figure*}

A blind line search for a narrow ($\sigma = 10$ eV) Gaussian line with free normalization and centroid energy applied to the baseline model was performed in the Fe K band between $5-11$ keV (rest-frame) for each epoch of observation, using a uniform step in energy $\Delta E=25$ eV \citep[e.g.,][]{2006MNRAS.366..115M}.
Results are shown in Fig.~\ref{FigLINES}, where the $\Delta\chi^2$ contours correspond to 68\%, 90\%, 99\%, and 99.9\% confidence level (from the outermost to the innermost, corresponding to $1,1.6,2.6,$ and $3.3$ standard deviations $\sigma$) in the centroid energy-normalization parameter space in the rest frame of PG 1126-041.  The three dashed vertical lines mark the rest-frame energy of Fe~I, Fe~XXV, and Fe~XXVI K$\alpha$ transitions.

Absorption features with significance $>99\%$ are visible in most of the spectra, with energies either between $E=7-7.5$ keV or at $E>9$ keV.
The small panels below each contour plot show the observed pn spectra as filled circles and the background as filled areas. During the 2004, 2009, and 2015 observations there was a strong background emission line at 9 keV (likely due to the Ni, Cu, and Zn in the detector), which is causing a spurious absorption feature once the background is subtracted from the source spectrum. Another emission line at $\sim 10$ keV is present in the 2009 background spectrum. We conclude that the highest-energy absorption features shown in the $\Delta\chi^2$ contour plots during the three epochs of observation 2004, 2009, and 2015 are not intrinsic to PG 1126-041 but are an effect of background subtraction. On the contrary, the absorption features between $7-7.5$ keV are confirmed to be intrinsic to the source at $>99\%$ confidence level in 5 out of 8 observations. Emission features at $>99\%$ confidence level are also present in half of the observations.

While the emission features centroid energy is compatible with rest-frame neutral iron emission or even redshifted emission, most of the absorption features are blueshifted, indicating that outflowing matter along the line of sight is present in most of the observations of PG 1126-041.
The absorption features are too blueshifted to be associated with lowly-ionized Fe, and the most conservative identification in terms of derived outflowing velocity is with the highly ionized Fe~XXV or Fe~XXVI K$\alpha$ and/or K$\beta$ transitions \citep[see, e.g., the discussion in Sect. 4.1 of][]{2010A&A...521A..57T}.
We attempt to model these residuals with two models: a phenomenological model in Sect.~\ref{sec:2009xstar}, and a physical accretion disk wind model in Sect. \ref{sec:all}.

\subsection{A phenomenological model\label{sec:2009xstar}}

Residuals larger than $3\sigma$ above and below the baseline model are present in the spectra of PG 1126-041 at $E > 4$ keV.
The negative residuals indicate the presence of absorbing gas of even higher ionization state than the $\log\xi\sim 2$ partially covering absorber; a result already found by  \citetalias{2011A&A...536A..49G}.
We generated a second absorption table with \texttt{XSTAR}, expanding the parameter space toward larger column densities and ionization parameters: the first was sampled at 5 points between $N_H=5\times 10^{22}$ and $N_H=10^{24}$ cm$^{-2}$, while the latter was sampled at 5 points between $\log\xi=2.5$ and $\log\xi=5$. Based on a fit of the high S/N 2009 dataset with \texttt{warmabs}, the turbulent velocity for this grid was set to 5000 km~s$^{-1}$ and the name of the grid to \texttt{xstar5000}.

\begin{table*}
\footnotesize{
\caption{\label{table:fitALLGAUSS}Spectral fit results for the phenomenological model \texttt{[(partcov*xstar500)*(xstar5000*pow + zgauss1 + zgauss2)]}.}
\centering
\begin{tabular}{lllllllll}
\hline\hline
 & 2004 & 2008A & 2008B & 2009 & 2014A & 2014B & 2014C & 2015 \\
 \hline
& \multicolumn{8}{c}{\textit{Power law emission:} \texttt{pow}}\\
$\Gamma$  & $2.02^{+0.07}_{-0.01}$ & $1.89^{+0.02}_{-0.02}$ & $2.03^{+0.01}_{-0.03}$ & $1.96^{+0.01}_{-0.01}$ & $2.20^{+0.03}_{-0.02}$ & $1.98^{+0.01}_{-0.02}$ & $1.89^{+0.01}_{-0.01}$ & $1.91^{+0.01}_{-0.01}$\\
$N_{1\,keV}/10^{-4}$  & $6.6^{+0.1}_{-0.1}$ & $8.1^{+0.4}_{-0.3} $ & $12.7^{+0.6}_{-0.1} $ & $5.5^{+0.1}_{-0.1} $ & $
3.0^{+0.1}_{-0.2} $ & $7.4^{+0.2}_{-0.1} $ & $8.4^{+0.2}_{-0.1} $ & $12.2^{+0.2}_{-0.2} $\\
  \multicolumn{9}{c}{ }\\
&\multicolumn{8}{c}{\textit{Partially covering absorber:} \texttt{partcov*xstar500}}\\
$N_H/10^{22}$   & $13.2^{+0.5}_{-0.2}$ & $8.7^{+0.2}_{-0.1}$ & $6.2^{+0.1}_{-0.1}$ & $13.6^{+0.1}_{-0.1}$ & $18.7^{+0.4}_{-0.3}$ & $14.5^{+0.1}_{-0.3}$ & $17.6^{+0.2}_{-0.2}$ & $11.9^{+0.2}_{-0.2}$ \\
$\log\xi$   & $1.91^{+0.01}_{-0.01}$ & $1.92^{+0.01}_{-0.01}$ & $1.71^{+0.02}_{-0.01}$ & $1.96^{+0.01}_{-0.01}$ & $2.06^{+0.01}_{-0.01}$ & $1.93^{+0.01}_{-0.01}$ & $2.02^{+0.01}_{-0.01}$ & $1.93^{+0.01}_{-0.01}$ \\
$C_f$ (\%) & $95.8^{+0.1}_{-0.1}$ & $96.3^{+0.4}_{-0.5}$ & $97.2^{+0.2}_{-0.5}$ & $94.6^{+0.1}_{-0.2}$ & $90.1^{+0.5}_{-0.2}$ & $96.2^{+0.2}_{-0.2}$ & $93.2^{+0.2}_{-0.3}$ & $95.7^{+0.2}_{-0.2}$\\
  \multicolumn{9}{c}{ }\\
&\multicolumn{8}{c}{\textit{Highly ionized absorber:} \texttt{xstar5000}}\\
$N_H/10^{22}$   & $32^{+26}_{-7}$ & $60^{+15}_{-25}$ & $15^{+3}_{-9} $ & $38^{+2}_{-3}$ & $60^{+5}_{-15}$ & $26^{+4}_{-12}$ & $76^{+15}_{-3}$ & $71^{+12}_{-6}$ \\
$\log\xi$   & \multicolumn{8}{c}{$3.55^{+0.01}_{-0.03}$} \\
$\upsilon_{out}/c$   &  $-0.045^{+0.006}_{-0.018}$ & $-0.063^{+0.012}_{-0.062}$ & $-0.11^{+0.02}_{-0.01}$ & $-0.063^{+0.007}_{-0.005}$ & $-0.06^{+0.03}_{-0.02}$  & $-0.06^{+0.02}_{-0.02}$ &  $-0.06^{+0.02}_{-0.01}$ & $-0.050^{+0.006}_{-0.011}$ \\
\multicolumn{9}{c}{ }\\
&\multicolumn{8}{c}{\textit{Gaussian emission line:} \texttt{zgauss1}} \\
$E$ (keV) & \multicolumn{8}{c}{$\equiv 6.4$ }  \\
$\sigma$ (eV)   & \multicolumn{8}{c}{$150^{+70}_{-50}$} \\
$I/(10^{-6})$  & $1.2^{+0.7}_{-0.6}$ & $<3.6  $ & $<3.2 $ & $1.8^{+0.5}_{-0.4}$ & $3.1^{+0.7}_{-0.6}$ & $<2.2$ & $3.8^{+1.0}_{-0.8} $ & $3.4^{+1.3}_{-1.1}$\\
EW  & $65^{+35}_{-35}$ & $<200 $ & $<150 $ & $95^{+30}_{-20}$ & $510^{+80}_{-70}$ & $<140 $ & $230^{+60}_{-40}$ & $140^{+50}_{-40}$\\
\multicolumn{9}{c}{ }\\
&\multicolumn{8}{c}{\textit{Gaussian emission line:} \texttt{zgauss2}} \\
$E$ (keV) & \multicolumn{8}{c}{$5.28^{+0.06}_{-0.07}$ }  \\
$\sigma$ (eV)   & \multicolumn{8}{c}{$1100^{+100}_{-100}$} \\
$I/(10^{-5})$  & $1.5^{+0.4}_{-0.9}$ & $1.4^{+0.6}_{-0.8}$ & $1.7^{+0.8}_{-1.1}$ & $1.7^{+0.1}_{-0.2}$ & $0.6^{+0.2}_{-0.3}$ & $1.7^{+0.4}_{-0.5}$ & $1.3^{+0.2}_{-0.4}$ & $0.7^{+0.5}_{-0.5}$\\
EW  & $640^{+120}_{-160}$ & $460^{+220}_{-260} $ & $350^{+160}_{-250}$ & $820^{+70}_{-120}$ & $820^{+250}_{-370}$ & $640^{+110}_{-200}$ & $690^{+140}_{-210}$ & $220^{+140}_{-170}$\\
\multicolumn{9}{c}{ }\\
&\multicolumn{8}{c}{\textit{MOS cross-calibration constant}} \\
$C$ & $1.18^{+0.02}_{-0.02}$ & $1.07^{+0.03}_{-0.03}$ & $1.10^{+0.02}_{-0.04}$ & $1.06^{+0.01}_{-0.01}$ & $1.10^{+0.02}_{-0.02}$ & $1.06^{+0.02}_{-0.02}$ & $1.00^{+0.02}_{-0.02}$ & $1.08^{+0.02}_{-0.02}$\\
\multicolumn{9}{c}{ }\\
\hline
\textit{Fit statistics} &  & \multicolumn{6}{c}{$\chi^2/\nu=2176/1888$} \\
 &   & \multicolumn{6}{c}{$P_{null}= 4\times 10^{-6}$} \\
\hline
$f_{0.3-10}^{observed}$ & $1.27^{+0.04}_{-0.01}$ & $1.70^{+0.14}_{-0.02}$ & $3.05^{+0.06}_{-0.09}$ & $1.24^{+0.01}_{-0.02}$ & $0.49^{+0.01}_{-0.02}$ & $1.56^{+0.03}_{-0.03}$ & $1.10^{+0.01}_{-0.04}$ & $1.69^{+0.04}_{-0.02}$\\
$f_{0.3-10}^{unabsorbed}$ & $3.66^{+0.10}_{-0.03}$ & $5.00^{+0.40}_{-0.07}$ & $7.46^{+0.15}_{-0.23}$ & $3.89^{+0.02}_{-0.07}$ & $2.81^{+0.05}_{-0.12}$ & $4.46^{+0.08}_{-0.08}$  & $3.62^{+0.04}_{-0.11}$ & $4.65^{+0.13}_{-0.06}$\\
$L_{2-10}^{unabsorbed}$ & $1.66^{+0.05}_{-0.01}$ & $2.15^{+0.17}_{-0.03}$ & $3.23^{+0.06}_{-0.10}$ & $1.62^{+0.01}_{-0.02}$ & $0.74^{+0.01}_{-0.03}$ & $2.11^{+0.04}_{-0.03}$  &  $1.54^{+0.02}_{-0.04}$ & $2.13^{+0.06}_{-0.02}$\\
\hline
\end{tabular}
\tablefoot{Units: power law normalisation $N_{1\,keV}$ [$10^{-4}$ photons keV$^{-1}$ cm$^{-2}$ s$^{-1}$]; column density $N_H$ [$10^{22}$ cm$^{-2}$]; ionization parameter $\xi$ [erg cm s$^{-1}$]; outflowing velocity $\upsilon_{out}$ [$c$]; Gaussian emission line centroid energy $E$ [keV], width $\sigma$ [eV], intensity $I$ [$10^{-5}$ photons cm$^{-2}$ s$^{-1}$], equivalent width EW [eV]; flux $f$ [$10^{-12}$ erg s$^{-1}$ cm$^{-2}$];  luminosity $L$ [$10^{43}$ erg s$^{-1}$]. The parameters kept tied between different epochs are the highly ionized absorber ionization parameter and the Gaussian emission line widths and centroid energies. The unabsorbed flux is corrected for both intrinsic and Galactic absorption, as is the luminosity. }}
\end{table*}

The baseline model with the addition of the highly ionized absorber \texttt{[(partcov*xstar500)*(xstar5000*pow)]} was fit with the \texttt{xstar5000} ionization parameter tied between epochs,  the column density and the velocity shift free to vary, and  all the baseline components free to vary between epochs.
The fit statistic is $\chi^2/\nu=2370/1906$.
The ionization parameter is $\log\xi\sim 3.5$, the column density is in the range between $3-5\times 10^{23}$ cm$^{-2}$ and the velocity shift is $- (0.045-0.11)\,c$. The velocity measured with \texttt{xstar5000} is a lower limit on the actual velocity of the wind because we only can measure the velocity component projected along our line of sight.

Despite the improvement of the fit statistic by $\Delta\chi^2/\Delta\nu=409/17$ with respect to the baseline model, the positive residuals larger than $3\sigma$ at $E > 4$ keV are not reproduced by the model.
We added to the model a  Gaussian emission line at the redshift of the source (\texttt{zgauss} in \texttt{Xspec}) with the centroid energy fixed to 6.4 keV, corresponding to the Fe I K$\alpha$ transition. The width of the emission line was kept constant between epochs, while the normalization of the line was left free to vary. We tested the two scenarios where the emission line is affected or not by the partially covering absorber and found better fit statistics in the first case ($\Delta\chi^2/\Delta\nu=78/9$ compared to $\Delta\chi^2/\Delta\nu=56/9$).
The line is narrow ($\sigma = 150^{+70}_{-50}$ eV) and has an average equivalent width $\langle\rm{EW}\rangle\sim 200$ eV, with a minimum during 2008A, 2008B, and 2014B (when only upper limits could be placed) and a maximum during 2014A (EW$\sim 500$ eV).
The fit statistic for the model \texttt{[(partcov*xstar500)*(xstar5000*pow + zgauss1)]} is $\chi^2/\nu=2292/1898$.

In order to account for the residuals observed at energies lower than 6.4 keV, we added a second Gaussian emission line with centroid energy and width free to vary but kept linked between epochs. The fit statistic improves by $\Delta\chi^2/\Delta\nu=116/10$ (F-test probability $> 99.999\%$) for a broad ($\sigma = 1100\pm{100}$ eV) emission line centered at $E = 5.3\pm{0.1}$ keV. The line has an average equivalent width $\langle\rm{EW}\rangle\sim 580$ eV. We consider this component not necessarily a physical emission line but a way to model a continuum more complex than a power law.
The fit statistic for this  model \texttt{[(partcov*xstar500)*(xstar5000*pow + zgauss1 + zgauss2)]} is $\chi^2/\nu=2176/1888$.
The intrinsic power law emission has an average $\langle \Gamma\rangle = 1.98$, and carries an average $2-10$ keV luminosity (corrected for absorption) $\langle L_{2-10}\rangle \sim 1.9\times 10^{43}$ erg s$^{-1}$, with a minimum value of $\sim 7\times 10^{42}$ erg s$^{-1}$ during 2014A and a maximum value of $\sim 3\times 10^{43}$ erg s$^{-1}$ during 2008B.
Spectral parameters are reported in Table~\ref{table:fitALLGAUSS}, while spectral residuals for this phenomenological model are shown in the left column of Fig. \ref{FIG_RES2}.

 \subsection{Disk wind model\label{sec:all}}

The positive and negative residuals with respect to the baseline model visible in the spectra of PG 1126-041 (see Fig.~\ref{FigLINES}) might be explained by scattering and absorption of photons into the accretion disk wind, a scenario introduced by \citetalias{2008MNRAS.388..611S,2010MNRAS.404.1369S}.
We tested this scenario using the extended grid \texttt{fast32}, which is a large collection of spectral simulations of the \citetalias{2010MNRAS.404.1369S} disk wind model extended by \citet{2022MNRAS.515.6172M}.
The wind is assumed to be smooth and stationary, with a biconical axisymmetric geometry and a  wind opening angle 45$^{\circ}$ with respect to the polar axis \citep[Fig. 1 of][]{2022MNRAS.515.6172M}.
The X-ray source of continuum emission is located at the origin of the coordinate system and has a size of $6r_g$ ($r_g\equiv GM_{BH}/c^2$ is the gravitational radius, where $G$ is the gravitational constant and $c$ is the speed of light), and the wind inner launching radius is $R_{min}=32\,r_g$.
While in the \texttt{fast32} grid the slope of the ionizing continuum is a variable parameter, in the extended \texttt{fast32} grid used in this work it has been fixed to save computational time to a power law with $\Gamma=2$.
The velocity structure of the wind is assumed to follow a simple $\beta-$law, $\upsilon(R)\propto\upsilon_{\infty}(1-R_{min}/R)^{\beta}$ with $\beta=1$, and special relativity effects are taken into account.
The ionization state of the wind is computed self-consistently, where K-, L-, and M-shell transitions of the most abundant cosmic ions are taken into account, and absorption, scattering, and reflection of photons into the wind are computed through Monte Carlo radiative transfer methods  \citep[see][for details]{2022MNRAS.515.6172M}.
The extended \texttt{fast32} disk wind model free parameters are: the mass outflow rate normalized to the Eddington value $\mw=\dot{M}_{w}/\dot{M}_{Edd}$; the ratio of $2-10$ keV  luminosity over the Eddington luminosity $L_X/L_{Edd}$; the cosine $\mu$ of the inclination angle $\theta$ between the line of sight and the polar axis; and the ratio between the terminal velocity on the wind streamline and the escape velocity, $f_{\upsilon} = \upsilon_{\infty}/\upsilon_{esc}$, thus from equating the observed velocity with the escape velocity at a radius of $32\, r_g$, for $f_{\upsilon}=1$ then $\upsilon_{\infty}=0.25\,c$.

We added  the grid \texttt{fast32} to the baseline model (fit statistics $\chi^2/\nu=2779/1923$), resulting in the model \texttt{[(partcov*xstar500)*(fast32*pow)].}
The \texttt{fast32} parameters $\mw$ and $\mu$ were kept constant during the different epochs, while $L_X/L_{Edd}$ was allowed to scale proportionally to the $2-10$ keV flux (corrected for absorption) in each epoch.
We tested three different thicknesses of the wind, with outer launching radius $R_{max}/R_{min}=1.5,3,5$, and found the best representation of the data for the case $R_{max}/R_{min}=3$. In general, a larger wind thickness implies a smaller density for a given column density and this is compensated by a larger $\mw$ \citep{2022MNRAS.515.6172M}. We found a fit statistic $\chi^2/\nu=2327/1912$ for a large inclination angle of our line of sight with respect to the biconical wind polar axis, $\theta\sim 82^{\circ}$, meaning that we are looking through the base of the wind.
The wind terminal velocity takes into account the geometry of the system, and is found to be $\langle \upsilon_{\infty}\rangle = -0.22c$, a factor almost $4\times$ larger than the velocity along the line of sight measured with the 1D \texttt{XSTAR} model in the previous section.
The mass outflow rate normalized to Eddington is $\mw=0.24$, while the $2-10$ keV ionizing luminosity ranges from 0.1\% to 0.47\% of Eddington.

The partial covering absorber parameters are consistent with those found with the phenomenological model in the previous Sect.~\ref{sec:2009xstar}.
In fact, modeling the spectra with the \texttt{fast32} wind component takes away the need for both the \texttt{xstar5000} component and the broad emission line at $E\sim 5.3$ keV and does not strongly change the covering fraction, column density, or ionization state of the partially covering absorber.
The same conclusion holds when the \texttt{fast32} model is replaced altogether by the relativistically blurred reflection model \texttt{relxill} \citep{2014ApJ...782...76G}, confirming the ability of the baseline model to reproduce the broadband X-ray spectral shape of PG 1126-041. In this case, the excess of photons in the energy range $4-6$ keV is well modeled, but the negative residuals in the Fe K band are not taken into account, overall giving a much worse fit statistics than the \texttt{fast32} model ($\chi^2/\nu=3120/1910$).

The narrow emission line at 6.4 keV is instead not reproduced by \texttt{fast32}, and can be either modeled with a phenomenological Gaussian emission line or with a self-consistent reflection model.
In the former case, the improvement in fit statistic is $\Delta\chi^2/\Delta\nu=33/9$ (F-test probability $>99.7\%$), with the line detected during the 2009, 2014A, and 2014C epochs. In the latter case, we used the \texttt{xillver} reflection model \citep{2010ApJ...718..695G,2013ApJ...768..146G} fixing the photon index $\Gamma=2$, the density of the reflecting material $n=10^{15}$ cm$^{-3}$, and the metal abundance to the solar value. The reflection component needs to be absorbed by the partially covering gas, and we obtain a $\Delta\chi^2/\Delta\nu=22/9$ (F-test probability $>94\%$) for an inclination angle $<45^{\circ}$ and a low ionization parameter $\log\xi < 0.15$. In both the Gaussian and the reflection scenarios, the partially covering absorber and the disk wind model parameters are not affected by the inclusion of the Fe K emission components.

  \begin{figure}[htb!]
  \centering
  \includegraphics[width=8.8cm]{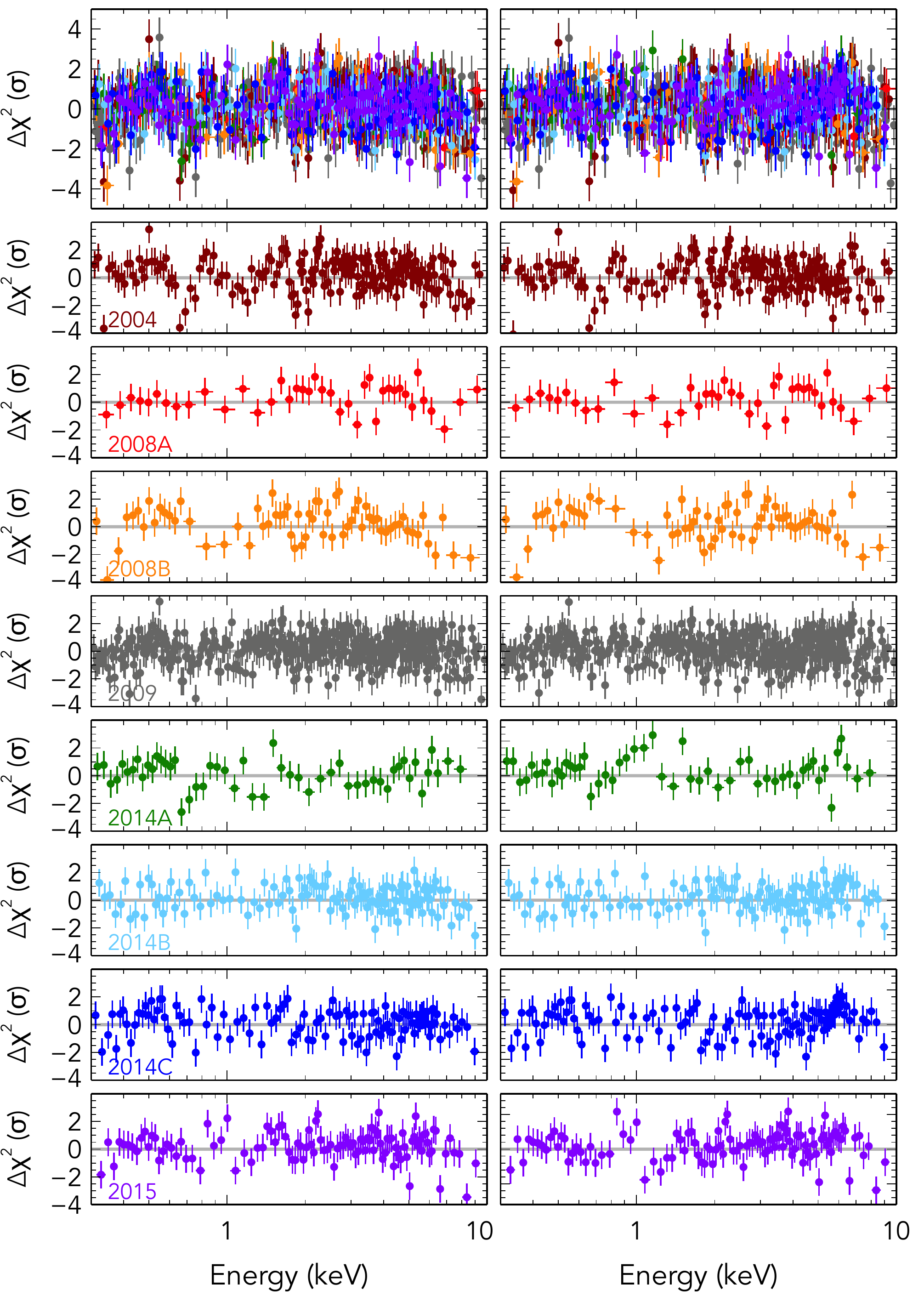}
\caption{EPIC-pn spectral residuals of PG 1126-041 to the phenomenological model  \texttt{[(partcov*xstar500)*(xstar5000*pow + zgauss1 + zgauss2)]} (left column) and to the disk wind model \texttt{[(partcov*xstar500)*(fast32*pow)]} (right column).
The top panel reports the residuals of all eight epochs together, while the smaller individual panels correspond to different epochs of observation, as listed in Table~\ref{T_OBS}.
   \label{FIG_RES2}}
    \end{figure}

The intrinsic power law continuum has a photon index $\langle\Gamma\rangle=1.85$.
The power law normalization best fit value, which depends on the number of photons lost from our line of sight due to absorption or scattering, is much larger when \texttt{fast32} is used compared to the phenomenological modeling.
In fact, \texttt{fast32} includes the effect of electron scattering, and for a given optical depth $\tau=\sigma N_H$ takes into account the decrease in amplitude of the observed emission, which is $\propto e^{-\tau}$.
This effect is not included in the simple \texttt{xstar5000} model, and as a consequence, the measured intrinsic luminosity of the source is underestimated.
The average $2-10$ keV luminosity corrected for absorption computed with the disk wind model is $\langle L_{2-10}\rangle = 5.7 \times 10^{43}$ erg s$^{-1}$, with a minimum $\sim 2\times 10^{43}$ erg s$^{-1}$ during 2014A and a maximum $\sim 9\times 10^{43}$ erg s$^{-1}$ during 2008B, a factor almost $3\times$ higher compared to the phenomenological modeling.

We untied the value of the mass outflow rate between epochs, but we did not obtain a significant improvement of the fit statistic.
Spectral parameters of the \texttt{[(partcov*xstar500)*(fast32*pow)]} model along with their errors and the associated fluxes and luminosities, are reported in Table~\ref{table_all_dw}. Spectral residuals are shown in the right column of Fig. \ref{FIG_RES2}.

 \begin{table*}
\footnotesize{
\caption{\label{table_all_dw}Spectral fit results for the \texttt{[(partcov*xstar500)*(fast32*pow)]} model.  The accretion disk wind  inclination angle and mass outflow rate are kept constant between epochs, while the ionizing hard X-ray luminosity $L_x/L_{Edd}$ is tied to the unabsorbed $2-10$ keV flux for each epoch.}
\centering
\begin{tabular}{lllllllll}
\hline\hline
 & 2004 & 2008A & 2008B & 2009 & 2014A & 2014B & 2014C & 2015 \\
 \hline
& \multicolumn{8}{c}{\textit{Power law emission:} \texttt{pow}}\\
$\Gamma$ & $1.90^{+0.01}_{-0.02}$ & $1.80^{+0.01}_{-0.05}$ & $1.78^{+0.01}_{-0.02}$ & $1.85^{+0.01}_{-0.01}$ & $2.09^{+0.01}_{-0.03}$ & $1.79^{+0.01}_{-0.01}$ & $1.87^{+0.01}_{-0.01}$ & $1.75^{+0.03}_{-0.01}$   \\
$N_{1\,keV}/10^{-4}$  & $36.7^{+0.4}_{-0.4} $ & $32.9^{+0.7}_{-0.6}$ & $50.4^{+1.3}_{-0.6} $ & $33.5^{+0.2}_{-0.3} $ & $24.6^{+0.4}_{-0.6} $ & $35.8^{+0.6}_{-0.4}$ & $31.6^{+0.6}_{-0.3} $ & $35.9^{+0.3}_{-0.8} $\\
  \multicolumn{9}{c}{ }\\
&\multicolumn{8}{c}{\textit{Partially covering absorber:} \texttt{partcov*xstar500}}\\
$N_H/10^{22}$   & $13.5^{+0.2}_{-0.2}$ & $7.0^{+0.2}_{-0.1}$ & $5.2^{+0.1}_{-0.1}$ & $14.9^{+0.1}_{-0.1}$ & $20.7^{+0.2}_{-0.7}$ & $14.0^{+0.3}_{-0.1}$ & $18.2^{+0.1}_{-0.3}$ & $12.1^{+0.2}_{-0.2}$ \\
$\log\xi$   & $1.92^{+0.01}_{-0.01}$ & $1.85^{+0.03}_{-0.02}$ & $1.78^{+0.04}_{-0.01}$ & $1.98^{+0.01}_{-0.01}$ & $2.05^{+0.02}_{-0.01}$ & $1.95^{+0.01}_{-0.01}$ & $2.00^{+0.01}_{-0.01}$ & $1.95^{+0.01}_{-0.01}$ \\
$C_f$ (\%) & $95.9^{+0.2}_{-0.1}$ & $95.0^{+0.4}_{-0.7}$ & $99.6^{+0.2}_{-0.9}$ & $94.7^{+0.1}_{-0.2}$ & $89.6^{+0.4}_{-0.4}$ & $96.0^{+0.2}_{-0.3}$ & $94.2^{+0.1}_{-0.3}$ & $95.5^{+0.3}_{-0.3} $\\
  \multicolumn{9}{c}{ }\\
 & \multicolumn{8}{c}{\textit{Disk wind:}\texttt{fast32}}\\
$L_X/L_{Edd}$ (\%) &  $0.243^{+0.012}_{-0.003}$ & $0.311^{t}$ & $0.471^{t}$ & $0.243^{t}$ & $0.104^{t}$ & $0.311^{t}$ & $0.228^{t}$  & $0.323^{t}$  \\
$\upsilon_{\infty}/c$   &  $-0.203^{+0.009}_{-0.005}$ & $-0.22^{+0.04}_{-0.04}$ & $-0.17^{+0.02}_{-0.03}$ & $-0.235^{+0.003}_{-0.005}$ & $-0.25^{+0.01}_{-0.01}$ & $-0.222^{+0.015}_{-0.005}$ & $-0.247^{+0.014}_{-0.005}$  & $-0.222^{+0.005}_{-0.008}$
\\
$\mu$ & \multicolumn{8}{c}{$0.132^{+0.006}_{-0.002}$}\\
$\mw$  &  \multicolumn{8}{c}{$0.236^{+0.003}_{-0.008}$}\\
\multicolumn{9}{c}{ }\\
&\multicolumn{8}{c}{\textit{MOS cross-calibration constant}} \\
$C$ & $1.11\pm{0.02}$ & $1.05\pm{0.04}$ & $1.13^{+0.09}_{-0.02}$ & $1.07\pm{0.01}$ & $1.10^{+0.03}_{-0.05}$ & $1.05\pm{0.03}$ & $1.00\pm{0.03}$ & $1.08\pm{0.03}$\\
\multicolumn{9}{c}{ }\\
\hline
Fit statistics & $\chi^2/\nu$ & \multicolumn{6}{c}{2322/1912} \\
 & $P_{null}$ & \multicolumn{6}{c}{$3\times 10^{-10}$} \\
\hline
$f_{0.3-10}^{observed}$ & $1.24^{+0.01}_{-0.04}$ & $1.68^{+0.10}_{-0.04}$ & $2.99^{+0.09}_{-0.06}$ & $1.24^{+0.01}_{-0.02}$ & $0.51^{+0.04}_{-0.01}$ & $1.50^{+0.03}_{-0.02}$ & $1.04^{+0.02}_{-0.02}$ & $1.71^{+0.07}_{-0.02}$\\
$f_{0.3-10}^{unabsorbed}$ & $12.2^{+0.1}_{-0.4}$ & $12.4^{+0.7}_{-0.3}$ & $19.8^{+0.6}_{-0.4}$ & $11.4^{+0.1}_{-0.2}$ & $4.8^{+0.4}_{-0.1}$ & $13.7^{+0.4}_{-0.1}$  & $9.9^{+0.2}_{-0.2}$ & $11.3^{+0.5}_{-0.1}$\\
$L_{2-10}^{unabsorbed}$ & $5.7^{+0.1}_{-0.2}$ & $6.1^{+0.4}_{-0.1}$ & $9.0^{+0.2}_{-0.2}$ & $5.3^{+0.1}_{-0.1}$ & $2.2^{+0.1}_{-0.1}$ & $6.5^{+0.1}_{-0.1}$  &  $4.7^{+0.1}_{-0.1}$ & $6.0^{+0.3}_{-0.1}$\\
\hline
\end{tabular}
\tablefoot{Units: power law normalisation $N_{1\,keV}$ [$10^{-4}$ photons keV$^{-1}$ cm$^{-2}$ s$^{-1}$]; column density $N_H$ [$10^{22}$ cm$^{-2}$]; ionization parameter $\xi$ [erg cm s$^{-1}$]; terminal velocity $\upsilon_{\infty}$ [$c$]; flux $f$ [$10^{-12}$ erg s$^{-1}$ cm$^{-2}$];  luminosity $L$ [$10^{43}$ erg s$^{-1}$]. The unabsorbed flux is corrected for both intrinsic and Galactic absorption, as is the luminosity. }}
\end{table*}

\subsection{X-ray partially covering absorber variability: a Bayesian approach\label{sec:partcov}}

Independent of the modeling adopted to reproduce the broadband $0.3-10$ keV spectral shape of PG 1126-041, variability of the partially covering absorber is observed between all epochs of observation.
In order to constrain these variations taking into account possible degeneracies between spectral parameters, and thus test the robustness of our spectral fit results obtained with the $\chi^2$ statistic, we explored the parameter space of both the phenomenological model and the disk wind model using a Bayesian approach.
We used the Bayesian X-ray Analysis (\texttt{BXA}) v.4.0.6 \citep{2014A&A...564A.125B}, which is an interface between \texttt{Xspec} and the Bayesian inference package \texttt{Ultranest}, that uses the most advanced nested sampling algorithm in terms of robustness, correctness, and speed \citep{2021JOSS....6.3001B}.
The Bayesian approach allows us to scan the parameter space without losing information due to, e.g., spectral binning, without having to make assumptions on the connections between different model parameters, and without having to worry ending up in local statistical minima. The whole parameter space is scanned at once with nested sampling methods, with the likelihood function being marginalized with probability weights given by the prior probability density. We used a Poisson log-likelihood function and analyzed both the grouped and ungrouped spectra, finding consistent results.

We assumed no prior knowledge and fitted first the highest S/N, 2009 epoch, with uninformed priors on all the parameters. The power law photon index was allowed to range between $\Gamma = [1.5-2.3]$ and its normalization between $N_{1\,keV} =[3\times 10^{-4}-3\times 10^{-2}]$; the partial covering column density, ionization state, and covering fraction between $N_H = [3-30\times 10^{22}]$ cm$^{-2}$, $\log\xi = [1.5-2.5]$, and $C_f = [0.8-1.0]$.
In the phenomenological  model, the  \texttt{xstar5000} column density, ionization state, and outflow velocity were allowed to range between $N_H = [2-10\times 10^{23}]$ cm$^{-2}$, $\log\xi = [2.5-4]$, and $|\upsilon_{out}| = [0.001-0.2]c$. To speed up the computational time, when analyzing the data with the phenomenological model we did not include the emission lines (modeled with Gaussians in Sect.~\ref{sec:2009xstar}). The presence or absence of these high-energy ($E > 5$ keV) components has a negligible effect on the estimate of the properties of the X-ray partially covering absorber, that affects lower-energy photons.
In the disk wind model, the \texttt{fast32}  mass outflow rate, cosine of the inclination angle, terminal velocity, and X-ray ionizing luminosity were allowed to range between $\mw = [0.02-1.25]$,  $\mu=[0.05-0.9]$ (i.e., $\theta$ between $25^{\circ}$ and $87^{\circ}$), $\upsilon_{\infty}/c = [0-0.25]c$, and  $L_X/L_{Edd} = [0.026-2.5]$, i.e., the full range of parameters included in the grid.

Fig.~\ref{FIG1} shows that there are no dramatic variations either in flux or in the spectral shape of PG 1126-041 between the different epochs of \textit{XMM-Newton} observations. The median values of the posterior probability distributions of the parameters found for the 2009 epoch were therefore used as informed Gaussian priors when fitting the other epochs. All the epochs were fitted independent of each other.

Results are reported in the Appendix for the 2009 data, where we plot the one- and two-dimensional histograms of the marginal posterior probability distribution for each spectral parameter (corner plot) using the phenomenological model (Fig.~\ref{FIGbxa2009xstar}) and the disk wind model (Fig.~\ref{FIGbxa2009}). In the top right corner of these figures, we plot the posterior probability distributions of the theoretical model (top) and of the model convolved with the instrumental response, with the EPIC-pn data overplotted (bottom). Every solid line is a representation of the model that gives a solution drawn from the posterior probability distribution; therefore the darker and thicker the line, the more probable is the solution.

Most of the statistical solutions for the model parameters found with the $\chi^2$ minimization are close to the median of the posterior probability distribution found with the Bayesian analysis. The Bayesian parameter estimation scans the whole parameter space at once and is able to show clearly any interdependency between parameters as well as the presence of multiple minima in the $\chi^2$ space. The parameter interdependencies are visible in the corner plots as 2D histograms that strongly deviate from being symmetric, such as, e.g., the one for the power law normalization $N_{1\,keV}$ versus the column density $N_H$ of the highly ionized absorber in the phenomenological model, or the one for the wind inclination angle and terminal velocity in the disk wind modeling. In this latter case, a double minimum in the statistical space is revealed: the solution found with the $\chi^2$ minimization is not unique but is accompanied by another with lower inclination ($\theta\sim 60^{\circ}$) and smaller terminal velocity ($\upsilon_{\infty}\sim 0.13c$).

\begin{figure}[hbt!]
  \centering
  \includegraphics[width=7.cm]{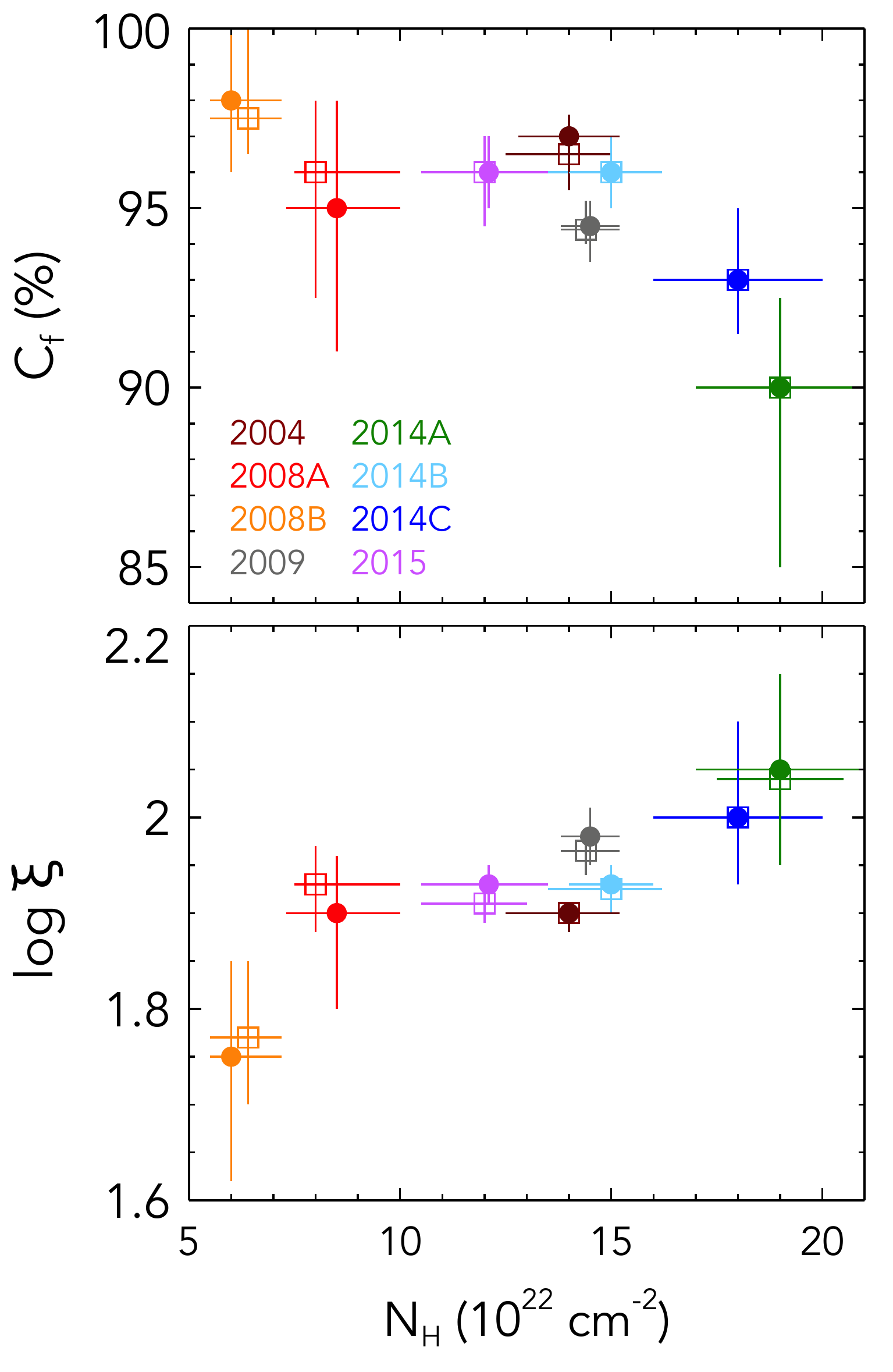}
   \caption{Column density of the partially covering X-ray absorber versus its covering fraction (top panel) and ionization parameter (bottom panel), derived from the Bayesian analysis of the EPIC-pn spectra in different epochs assuming the phenomenological model (empty squares) or the disk wind model (filled circles). Error bars represent $2\sigma$ deviations from the median posterior probability distribution of the parameters.}
   \label{FIG9}%
    \end{figure}

The X-ray partially covering absorber parameters, on the other hand, are well-constrained in all the epochs of observation.
Figure \ref{FIG9} shows the median values of the posterior probability distribution of the column density, the covering fraction,  and the ionization parameter of the X-ray partially covering absorber measured at different epochs. Here we plot with open squares the values referred to the phenomenological model and with filled circles those referred to the disk wind model; each error bar represents the $2\sigma$ equivalent probability, as derived from the 2D histograms of the posterior probability distribution.
The Bayesian analysis confirmed that there are variations in column density $>10\%$ between every consecutive observation, independent of the model adopted.
The average value $\langle N_H \rangle = 13.5 \times 10^{22}$ cm$^{-2}$ is much larger than the values measured for the X-ray warm absorbers and at the lower range of the columns measured in high velocity X-ray ultrafast outflows \citep[e.g.,][]{2021NatAs...5...13L}.
The moderate spectral resolution of the EPIC cameras
does not allow to measure the velocity of the partially covering absorber.
A minimum column density is measured in the 2008B observation, $N_H= 6.1\pm{0.5}\times 10^{22}$ cm$^{-2}$; and a maximum in the 2014A observation, with $N_H = 1.9\pm{0.1}\times 10^{22}$ cm$^{-2}$. The total variation in column density is by a factor larger than $3\times$.

Remarkably, there are  variations of $N_H$ on time scales as short as the separation between the 2014A, 2014B, and 2014C observations (panels d, e, and f in Fig. \ref{FIG1} and green, cyan, and blue points in Fig. \ref{FIG9}).
The column density decreases by $ >20\%$ $(\Delta N_H\sim 4\times 10^{22}$ cm$^{-2}$) during the 11 days elapsed between 2014A and 2014B and then increases again by $\sim 20\%$ during the 16 days elapsed between 2014B and 2014C $(\Delta N_H\sim 3\times 10^{22}$ cm$^{-2}$). Finally, there is a further drop in $N_H$ by $\sim 30\%$ between 2014C and 2015 $(\Delta N_H\sim 6\times 10^{22}$ cm$^{-2}$).

\section{UV C IV absorption variability \label{sec:XUV}}

      \begin{figure}[h!tb]
  \centering
    \includegraphics[width=8.3cm]{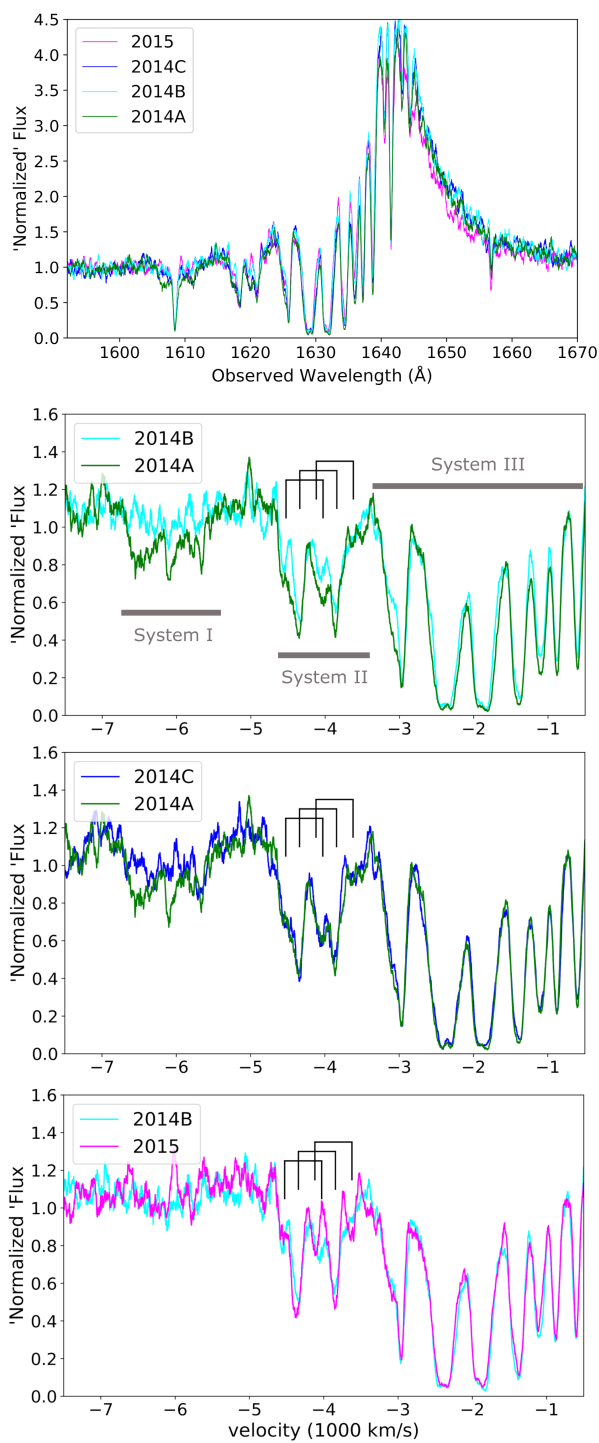}\caption{ \textit{HST}-COS spectra of the {\CIV} region of PG 1126-041 taken between 2014 and 2015.  Top panel: the four spectra are plotted overlapped after being normalized and matched at $1595-1605$ {\AA} to show both changes in emission and absorption. Three bottom panels: zoom on the $1600-1640\AA$ region, comparing different couples of epochs in velocity scale, after normalization using a 2nd-order polynomial fit. Three velocity systems are marked, as well as three likely {\CIV} doublets in system II.
The strong absorption line at 1608\r{A} is due to Galactic {\FeII} and was removed prior to the measurements, so it does not appear in the three bottom panels. All spectra in this figure have been smoothed with a boxcar filter with a width of 31 pixels (corresponding to less than 0.4 \AA\, or 75 km s$^{-1}$) for visualization purposes.}
   \label{FigUV2}%
    \end{figure}

\begin{table*}[ht!]
\caption{Log of the \textit{HST}-COS observations of PG 1126-041  \label{T_OBS_HST}}
\centering
\begin{small}
\begin{tabular}{ccccc}
\hline\hline
Name & Grating/Central Wavelength &  Start date  &  t$_{exp}$ &  $\Delta$t$_{HST-XMM}$ \\
(1) & (2) & (3) & (4) & (5) \\
\hline
 2014A & G130M/1327 {\AA} & 2014-06-01 03:05:15 & 1.580 & 0.13 \\
& G130M/1055 {\AA} & 2014-06-01 03:37:30 & 1.874 & 0.10 \\
& G160M/1577 {\AA} & 2014-06-01 05:01:22 & 1.079 & 0.05 \\
 2014B & G130M/1327 {\AA} & 2014-06-12 16:19:16 & 1.580 & 0.18 \\
    & G130M/1055 {\AA} & 2014-06-12 16:51:31 & 1.874 & 0.16 \\
    & G160M/1577 {\AA} & 2014-06-12 18:14:30 & 1.079 & 0.10 \\
 2014C & G130M/1327 {\AA} & 2014-06-28 11:25:07 & 1.580 & 0.30 \\
 & G130M/1055 {\AA}& 2014-06-28 11:57:22 & 1.874 & 0.30  \\
 & G160M/1577 {\AA}& 2014-06-28 13:21:29 & 1.079 & 0.22 \\
 2015 & G130M/1327 {\AA} & 2015-06-14 07:15:17 & 1.540 & 0.007 \\
 & G130M/1055 {\AA} & 2015-06-14 07:47:56  & 1.837 & 0.03 \\
 & G160M/1577 {\AA} & 2015-06-14 09:12:04  & 1.060 & 0.09 \\
\hline
\end{tabular}
\end{small}
\tablefoot{(1) Name used in the article; (2) Grating and the central wavelength at which it was aimed; (3) Starting date of observation (yyyy-mm-dd hh:mm:ss UTC); (4) Exposure time (ks); (5) Time separation between the start dates of the \textit{HST} and \textit{XMM-Newton} observations (days). In 2015, the \textit{HST} observations were taken during the \textit{XMM-Newton} observation (starting time: 2015-06-14 07:04:49 and exposure time 18 ks, see Table~\ref{T_OBS}).}
\end{table*}

  High-resolution ($R\sim 20,000$, dispersion 12.23 m\AA/pixel) UV observations were taken very close in time to the \textit{XMM-Newton} observations using the \textit{HST}-COS during the period 2014-2015; \textit{HST} programs 13429 and 13836 in cycles 21 and 22, respectively (PI: M. Giustini).
The average S/N per resolution element of the COS data in the {\CIV} region is $\sim$7.3.
Table \ref{T_OBS_HST} shows the gratings and central wavelengths used, the exposure times and the time difference between the starting time of the \textit{HST} and \textit{XMM-Newton} observations. All the \textit{HST} observations in 2015 were taken during the 18 ks \textit{XMM-Newton} observation. The 2014 \textit{HST} observations are very close in time to the corresponding \textit{XMM-Newton} observations, with the minimum time separation being 0.05 days and the maximum being 0.30 days.
Here we focus on the analysis of {\CIV} absorption, while the complete details and analysis of the COS spectrum will be reported in a companion article (Rodr\'iguez  Hidalgo  et  al.,  in prep.).

Figure \ref{FigUV2} (top panel) shows the region around the {\CIV} emission line in all epochs. The spectra in this figure have been smoothed with a boxcar filter 31 pixels wide.
In this top figure, the four spectra were only matched in the wavelength region $1595 - 1605 \,\AA$, so any observed variability is due to both changes in {\CIV} emission and {\CIV} absorption and not to changes in the continuum emission level.
{\CIV} absorption is present between $\sim 1605-1612 \,\AA$,  $\sim 1616-1622 \,\AA$,   and $\sim 1625-1637 \, \AA$.
In order to quantify the strength of these absorption features, we normalized the four spectra taking into account the emission and continuum around the absorption features. Similar to the work described in \citet{2013ApJ...775...14R}, we use second-order polynomial functions to mimic the slope of the blue side of the {\CIV} emission line. We fit the functions to four regions where absorption is not present in either spectra; the same regions were used for all epochs: $1602-1605\,\AA$, $1612.5-1615.5\,\AA$, $1623-1624\,\AA$, and $1637.8-1638.1\,\AA$, but different polynomial functions were used for each epoch (see Appendix \ref{AppendixB}). We then measured continuous absorption troughs between $1605-1637\,\AA$, defining absorption features as troughs present below $0.9\times$ of the continuum level for $>200$ {\kms}. The strong absorption line at $1608\,\AA$ is likely due to intervening Galactic {\FeII} absorption and is not variable; thus, it was removed from the spectra by fitting it with a single Gaussian. The three bottom panels of Fig.~\ref{FigUV2} show a zoom in of the region where outflowing {\CIV} was detected, comparing different epochs in pairs.

 For each absorption, feature we measure the maximum and minimum velocity ($\upsilon_{max}$ and $\upsilon_{min}$, respectively, where zero velocity lies at the quasar redshift $z=0.062$ and velocity limits are defined when the flux goes back up to $>0.9 \times$ the normalized flux), the equivalent width ($\rm EW$), and the maximum depth of the {\CIV} absorption troughs. The results are reported in Table \ref{UVabs}.
Errors in the measurements derive mostly from the
systematic uncertainty in the placement of the continuum fit. We follow a similar procedure as in \citet{2011MNRAS.411..247R}.

The {\CIV} absorption in PG 1126-041 displays three distinct absorption systems:

\begin{description}
\item[$\bullet$] System I: this is the system with the shortest wavelength ($1605 \,{\AA} \lesssim \lambda \lesssim 1612 \,{\AA}$), highest-velocity absorption. It shows the shallowest depth and the largest variability between all epochs of the three absorption systems.
It was not detected during 2015 and was only marginally detected during 2014B. Its maximum velocity is $\upsilon_{max}\sim -6700$ km s$^{-1}$ in epochs 2014A and 2014C and is slightly lower during 2014B ($\upsilon_{max}\sim -6530$ km s$^{-1}$). Its depth is maximum during 2014A and significantly decreases during the 2014B epoch, then increases again during 2014C; the EW variations follow a similar trend.\\
\item[$\bullet$] System II: shows absorption between $1617-1623\,\AA$ with a maximum velocity  of $\sim -4700$ km s$^{-1}$. It shows minimal variability in velocity but shows significant changes in EW and depth, being stronger during 2014A and 2014C and weaker during 2014B and 2015;\\
\item[$\bullet$] System III: this is the system with the longest wavelength and lowest-velocity absorption ($\upsilon_{max}\sim-3400$) {\kms}. It shows the largest EW and depth and almost no variability between all epochs.
\end{description}

The most extreme variability in the {\CIV} absorption features occurs between observations 2014A and 2014B (second panel of Fig.~\ref{FigUV2}), which are just separated by 11 days (10 days in the rest frame of PG 1126-041). System III overlaps mostly between all observations. However, systems I and II become shallower between these two observations; indeed, system I is present in 2014A at $1606-1607\,\AA$ but almost disappears in 2014B.
Fifteen days later, during the 2014C observation (dark blue in the third panel of Fig.~\ref{FigUV2}), system II returns to a strength and depth similar to the absorption in the 2014A spectrum, but system I remains as weak as in 2014B.
In 2015 (pink in the fourth panel of Fig.~\ref{FigUV2}), the absorption profiles are overall the weakest and resemble the ones observed in 2014B.

System II offers a remarkable pattern where the absorption returns to very similar depths at different observations: after weakening in 2014B, the absorption profile in the 2014C spectrum is very similar to the one in 2014A (shown in the third panel of Fig.~\ref{FigUV2}). Similarly, its absorption profile observed in 2015 is very similar to the one in 2014B (see the fourth panel of Fig.~\ref{FigUV2}).
System I and the highest-velocity part of the absorption complex of system II ($\upsilon_{max}\sim -4700$ km s$^{-1}$) show the largest variability in EW and depth. System I is also the weakest of the three absorption complexes.
This resembles what we observe in high-redshift BAL QSOs, where the strongest variability of the UV absorption troughs is observed in the weakest BALs and in those outflowing at the highest velocities \citep{2011MNRAS.413..908C}.

\begin{table}[ht!]
\caption{\textit{HST}-COS {\CIV} absorption measurements}
\begin{small}
\begin{tabular}{llllll}
\hline \hline
Epoch & System & $\upsilon_{max}$& $\upsilon_{min}$ & $\rm EW$ & Max  \\
 &   & (\kms) & (\kms) & (\kms) &  depth \\
\hline
2014A & I & -6710 & -5500 & 200 & 0.34 \\
& II & -4670 & -3400 &  380 &  0.62 \\
& III &  -3330 & -750 &  1510 &  0.98 \\
\hline
2014B & I & -6530 & -5900 & 40 & 0.14 \\
& II & -4640 &  -3490  & 290 &   0.55 \\
& III & -3370  & -750   & 1380  &  0.97  \\
\hline
2014C & I & -6730& -5560 & 90 & 0.20 \\
& II & -4770 & -3500 &  360 &  0.64  \\
& III & -3260  &  -750  & 1480   &  0.98  \\
\hline
2015 & II & -4620  &  -3750 & 260  &  0.60  \\
& III &-3470  &  -760 &  1320  &  0.96 \\
\hline
\end{tabular}
\label{UVabs}
\end{small}
\tablefoot{Errors in the $\upsilon_{max}$, $\upsilon_{min}$ and EW values derive mostly from the pseudo-continuum location of the polynomial fit. Typically, the errors are $\pm$20-100 {\kms} for the velocities, $\pm$20-50 {\kms} for EW, and $\pm$0.02 for the maximum depth. The absorption system I was not detected in 2015.
}
 \end{table}

\section{Summary and discussion \label{sec:DISCU}}

\subsection{The X-ray spectral properties of PG 1126-041\label{discu:xray}}

The nucleus of PG 1126-041 displays significant X-ray spectral variability, remarkably also between the three observations of 2014, which are separated by only about 10-15 days.
The observed X-ray flux is not dramatically varying between the eight different epochs of observation except for 2008B and 2014A, which stand out with the largest and lowest observed X-ray flux.
The average observed $0.3-10$ keV flux is  $1.5\pm{0.2}\times 10^{-12}$ erg cm$^{-2}$ s$^{-1}$; it is a factor $2\times$ higher during 2008B and a factor $3\times$ lower during 2014A.

The  spectral features that characterize the $0.3-10$ keV spectra of PG 1126-041 are:
\begin{description}
\item[$\bullet$] a broadband spectral curvature best reproduced by  ionized absorption partially covering the X-ray continuum emission source, which is modeled with a power law with photon index $\Gamma\sim 1.9$ (the ``baseline model'');\\
\item[$\bullet$] complexities at $E\sim 4-10$ keV, with both X-ray emission and absorption features not taken into account by the baseline model.
\end{description}

The ionized, partially covering absorbing gas is detected in all the  epochs of  observation, while emission and absorption features are detected in about a half of them.
 The ionization parameter of the partially covering absorber gives the maximum opacity to the continuum photons at $E\lesssim 3$ keV, where a substantial spectral curvature is predicted along with a deep and broad absorption trough between $0.6-1$ keV.  The spectral curvature is a deviation of the observed X-ray photon flux from the power law continuum emission model. It is due to the photoelectric cutoff and resonant absorption lines and moves to higher energies for larger column densities \citep[e.g.,][]{2001ApJS..133..221K}.
The broad absorption trough between $0.6-1$ keV is due to a large number of absorption lines and edges, and its depth is diluted by the presence of the non-negligible fraction of the power law emission that escapes unaffected from the partially covering absorber.
In particular, the decrease in covering fraction increases the soft X-ray flux at $E < 2$ keV, and this flux dilutes the broad absorption trough.

\begin{figure}[ht!]
  \centering
  \includegraphics[width=9cm]{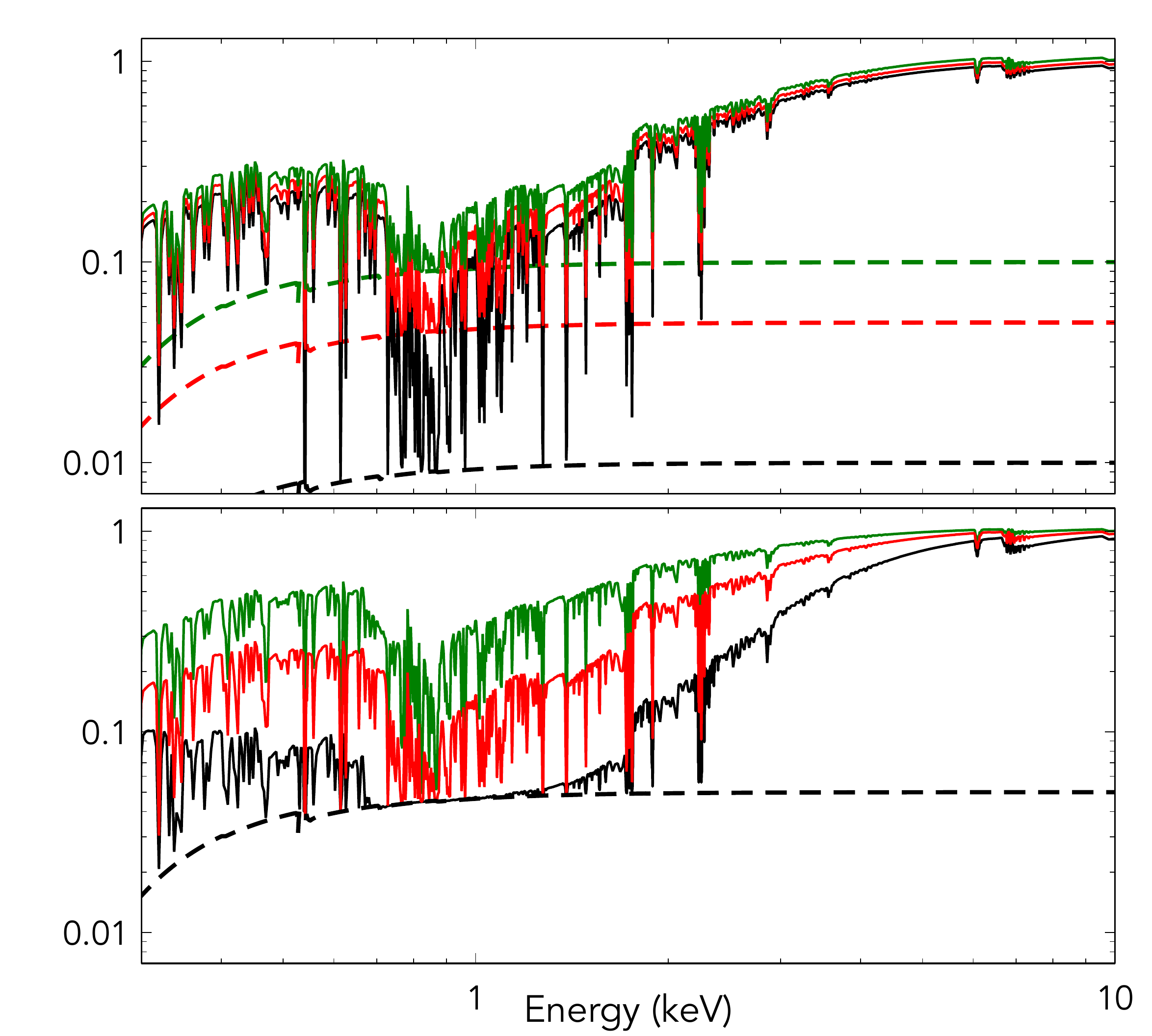}
   \caption{Theoretical models showing the effects of varying the covering fraction $C_f$ and the column density $N_H$ of a $\log\xi=2$ absorber partially covering a power law emission with $\Gamma=2$. Top panel: the column density is fixed to $N_H=10^{23}$ cm$^{-2}$ and the covering fraction increases from $C_f=90\%$ (green), to $95\%$ (red), to $99\%$ (black); the solid lines represent the total models, while the dashed lines represent the $(1-C_f)$ fractions of unabsorbed power law emission. Bottom panel: the covering fraction is fixed to   $C_f=95\%$ (the $5\%$ fraction of the unabsorbed power law is plotted with a dashed line), and the column density increases from $N_H=5\times 10^{22}$ cm$^{-2}$ (green), to $N_H= 10^{23}$ cm$^{-2}$ (red), to $N_H=2\times 10^{23}$ cm$^{-2}$ (black). The normalization of the y-axis is arbitrary.
   }
   \label{FIGPCOV}%
    \end{figure}

The different ways in which variations of column density or covering fraction affect the X-ray spectral shape can be seen in Fig.~\ref{FIGPCOV}. Here a power law emission with $\Gamma=2$ is partially covered by a layer of gas with $\log\xi=2$ and different column densities and covering fractions. The top panel shows the effect of variations of the covering fraction for a fixed column density $N_H=10^{23}$ cm$^{-2}$: at the highest $C_f$ value, the broad absorption trough between $\sim 0.6-1$ keV is very deep, while it gets shallower as $C_f$ decreases. The bottom panel shows the effects of variations in column density for a fixed $C_f=95\%$: the X-ray flux is absorbed at higher energies for larger column densities compared to lower column densities.
These theoretical expectations can be observed in the average flux state spectra (2009), compared to the lowest (2014A, panel d in Fig. \ref{FIG1}) and to the maximum flux state spectra (2008B, panel c in Fig. \ref{FIG1}). During 2008B, the covering fraction is maximum and the observed absorption trough between $\sim 0.6-1$ keV is the deepest; during 2014B, the covering fraction is minimum and the absorption trough is the shallowest. The spectral deviation from a simple $\Gamma=2$ power law happens at $E\sim 3$ keV during 2014A, when the column density is maximum, and at $E\sim 1.5$  keV during 2008B, when the column density is minimum; the 2009 case is intermediate between the two.
In summary, in the X-ray spectrum of PG 1126-041, the variations of the partially covering absorber covering fraction can be most appreciated in the soft X-ray band at $E \lesssim 1$ keV, while variations in its column density leave their signature at harder energies, $E \gtrsim 1$ keV.

The spectral complexities with respect to the baseline model
are visible in the right column of Fig. \ref{FIG_RESBASELINE}.
The results of a blind search for a Gaussian line either in emission or in absorption in the Fe K band of the eight epochs are shown in Fig. \ref{FigLINES}.
Residuals in absorption significant  at $>99.9\%$ confidence are present during 2004, 2009, and 2015. During 2009 and 2014C, there was a prominent and broad emission feature extending redward of 6 keV (rest-frame).
The negative residuals can be reproduced by a highly ionized outflowing absorber, the positive residuals by a  phenomenological Gaussian emission line (Sect. \ref{sec:2009xstar}), or all together by a disk wind model that takes into account both the X-ray photons absorbed along and those scattered back into the observer's line of sight (Sect. \ref{sec:all}).

Modeling the highly ionized absorber with the 1D photoionization code \texttt{XSTAR} gives a velocity blueshift $\langle\upsilon_{out}\rangle$ of about $- 0.06 c$. This is larger than the $10,000$ km s$^{-1}$ threshold used to define ultra-fast outflows \citep[UFOs,][]{2010A&A...521A..57T}; therefore, we will refer to this component as an UFO in the following.
The UFO velocity observed in PG 1126-041 is at the lower end of the UFO velocity distribution observed in local Seyferts, while the column density $\langle N_H\rangle \sim 5\times 10^{23}$ cm$^{-2}$ and the ionization parameter $\log\xi\sim 3.5$ are near the average \citep[][]{2011ApJ...742...44T,2013MNRAS.430...60G}.

The Gaussian emission line used to model the residuals in excess of the baseline model is very broad ($\sigma\sim 1$ keV) and centered at $E\sim 5.3$ keV in the source rest frame; therefore, it is unlikely that it corresponds to a physical individual emission feature.
An excess of counts at $E\sim 4-6$ keV  is often observed in the X-ray spectra of local AGN, and it is often interpreted within a relativistic reflection scenario or a complex partial covering absorption scenario \citep[e.g.,][]{2002MNRAS.331L..35F,2014PASJ...66..122M}. The two scenarios often give statistically equivalent fits to the data, for example in Mrk 335 \citep[e.g.,][]{2013MNRAS.428.1191G, 2008ApJ...681..982G,2015MNRAS.446..633G}, in 1H 0419-577 \citep[e.g.,][]{2005MNRAS.361..795F,2009ApJ...698...99T, 2014A&A...563A..95D}, in 1H 0707-495 \citep[e.g.,][]{2004MNRAS.353.1064G,2004PASJ...56L...9T,2012MNRAS.422.1914D}, and in PG 1535+547 \citep{2008A&A...483..137B}. In some cases, both relativistic reflection and complex absorption may contribute to shaping the X-ray spectra of AGN \citep[e.g.,][]{2009ApJ...696..160R,2012MNRAS.426.2522P,2021MNRAS.508.1798P}.

Both the photons in emission at $E\sim 4-6$ keV and the photons absorbed at $E>7$ keV might also be produced by scattering and absorption in an accretion disk wind \citep{2008MNRAS.388..611S,2010MNRAS.404.1369S, 2010MNRAS.408.1396S}.
It has been recently demonstrated by \citet{2022MNRAS.513..551P} that the emission expected from the accretion disk wind models is in fact spectrally degenerate with the relativistic reflection emission expected close to the SMBH, but the latter does not include the effects of absorption along the line of sight.
The accretion disk wind scenario has been successfully applied to reproduce AGN spectra at $E > 2$ keV \citep{2012ApJ...752...94T,2015MNRAS.446..663H,2016MNRAS.461.3954H} and also the broad-band X-ray spectra of individual AGN, i.e., PDS 456 \citep{2014ApJ...780...45R}, I Zw 1 \citep{2019ApJ...884...80R}, and MCG-03-58-007 \citep{2022ApJ...926..219B}.

\begin{figure}[htb!]
  \centering
  \includegraphics[width=8.5cm]{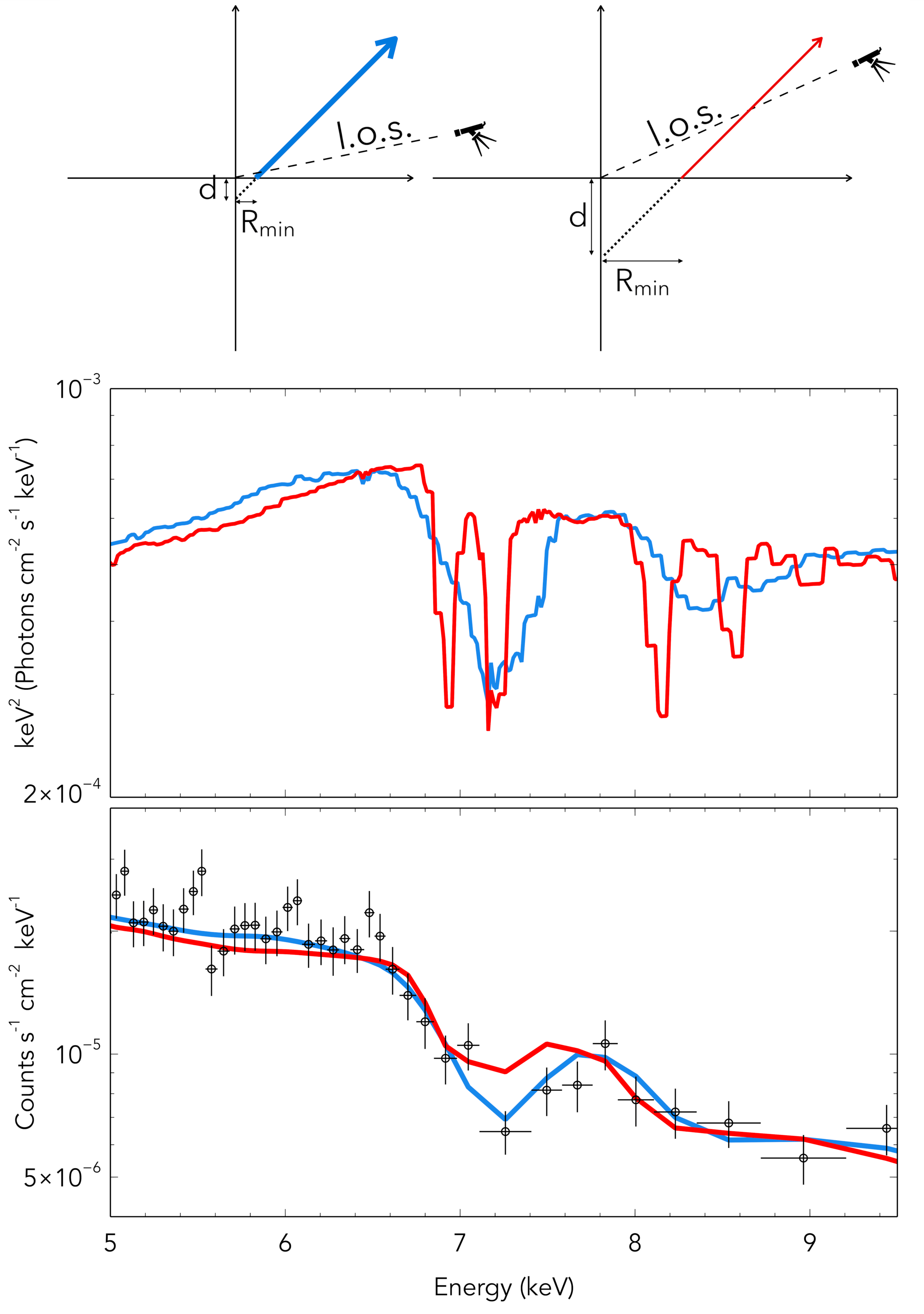}
   \caption{Two different geometries for the interpretation of the blueshifted absorption troughs in the Fe K band of PG 1126-041. A wind with a smaller $R_{in}$ (corresponding to $\upsilon_{\infty} = -0.25c$) and observed at a larger inclination angle ($\theta = 80^{\circ}$) is shown in blue, while a wind with a larger $R_{in}$ (corresponding to $\upsilon_{\infty} = -0.0625c$) observed at a smaller inclination angle ($\theta = 65^{\circ}$) is shown in red. Top panel: sketch of the geometry of the disk wind in the two cases (only $R_{in}$ has been sketched for simplicity), with streamlines represented by arrows with different thicknesses, proportional to the terminal velocity. Central panel: the two theoretical models. Bottom panel: the 2009 pn data (in gray) overplotted to the two models folded with the instrumental response.}
   \label{FIGLOS}%
    \end{figure}

The accretion disk wind model applied in this work is an extension of the spectral grids presented in \citet{2022MNRAS.515.6172M}.
The data require a very large inclination angle of the line of sight with respect to the biconical wind polar axis, $\theta\equiv\arccos\mu\sim 80^{\circ}$.
In this case, our line of sight goes through the wind base, and the observed spectrum is dominated by emission reprocessed by the wind.
Since the wind is assumed to be biconical and the inclination angle high, the terminal velocity of the wind is much larger than the observed projected velocity, $\upsilon_{out} \ll \upsilon_{\infty}\sim -0.2c$.
While, in principle, a wind launched at larger radii (and therefore with a lower terminal velocity) could lead to the same observed projected velocity for a lower inclination angle, the main constraint for such a large inclination angle comes from the very large depth of the Fe K absorption trough. This is shown in Fig.~\ref{FIGLOS}, with a simplified sketch of two geometries for the biconical accretion disk wind in PG 1126-041: the one corresponding to our best-fit scenario on the left in blue, and a wind launched at larger radii on the right, which has a terminal velocity close to the projected velocity  observed in PG 1126-041, in red. The middle panel shows the theoretical prediction for the Fe K band in the two scenarios: the wind launched farther out allows for a lower inclination angle (about $65^{\circ}$), almost parallel to the wind streamline, giving the same blueshift as the wind launched closer in and observed at a larger inclination angle (about $80^{\circ}$), but with much narrower absorption line profiles. The bottom panel shows the high S/N 2009 EPIC-pn data compared to the two models: the slower wind predicts a  shallower absorption trough in the Fe K band compared to the faster wind, which is therefore preferred by the data.
Future X-ray microcalorimeter observations of the Fe K band of PG 1126-041, with e.g., Resolve onboard \textit{XRISM} \citep{2020arXiv200304962X} or X-IFU onboard \textit{Athena} \citep{2018SPIE10699E..1GB,2023ExA....55..373B}, should allow a definitive distinction between the two scenarios.

Such a large inclination angle is so close to the base of the wind (or the atmosphere of the accretion disk) that very large column densities, highly Compton-thick, are expected by more realistic numerical simulations \citep{2010MNRAS.408.1396S}; for such equatorial lines of sight, the dusty torus at parsec-scales might also be intercepted, if present \citep[e.g.,][]{2017NatAs...1..679R};
however, the geometry of the torus is expected to vary with the evolution of AGN \citep[e.g.,][]{2012MNRAS.420..320H}.
Furthermore, the biconical geometry of the wind considered here is simple, and the physics ignores the effects of, for example, gas pressure and magnetic fields. Therefore, the results that we present should be taken as indicative and not as absolute measurements. It is possible that future hydrodynamical simulations could predict such large velocities and absorption depths for winds launched at larger radii than those considered here and/or observed at lower inclination angles.
In any case, the line of sight toward the nucleus of PG 1126-041 is likely intercepting the wind at every epoch of observation.

The intrinsic power law photon index is found to be $\Gamma\sim 1.8$, slightly flatter than the $\Gamma=2$ assumed in both the \texttt{XSTAR} and \texttt{fast32} model calculations.
The main effect of increasing the input $\Gamma$ in the calculation of the \texttt{fast32} model is to produce deeper absorption troughs in the Fe K band \citep{2022MNRAS.515.6172M}, thanks to a smaller number of hard X-ray photons able to over-ionize the iron atoms of the wind. Therefore, by assuming a slightly higher $\Gamma$ in our calculations, we might have underestimated the amount of matter necessary to produce the deep absorption trough, which in the model is parametrized by $\mw$.
The current uncertainties on the value of $\Gamma$ should be significantly reduced in the near future, thanks to approved broadband X-ray observations with \textit{XMM-Newton} + \textit{NuSTAR} (PI: J. N. Reeves) that should reveal the intrinsic continuum slope of PG 1126-041.

Overall, the disk wind model reproduces well the X-ray spectra of PG 1126-041 if highly variable, massive clumps, which are represented here by the X-ray partial covering absorber, are included in the modeling.

\subsection{The X-ray/UV connection\label{discu:xuv} }

\begin{figure*}[htb!]
  \centering
  \includegraphics[width=16cm]{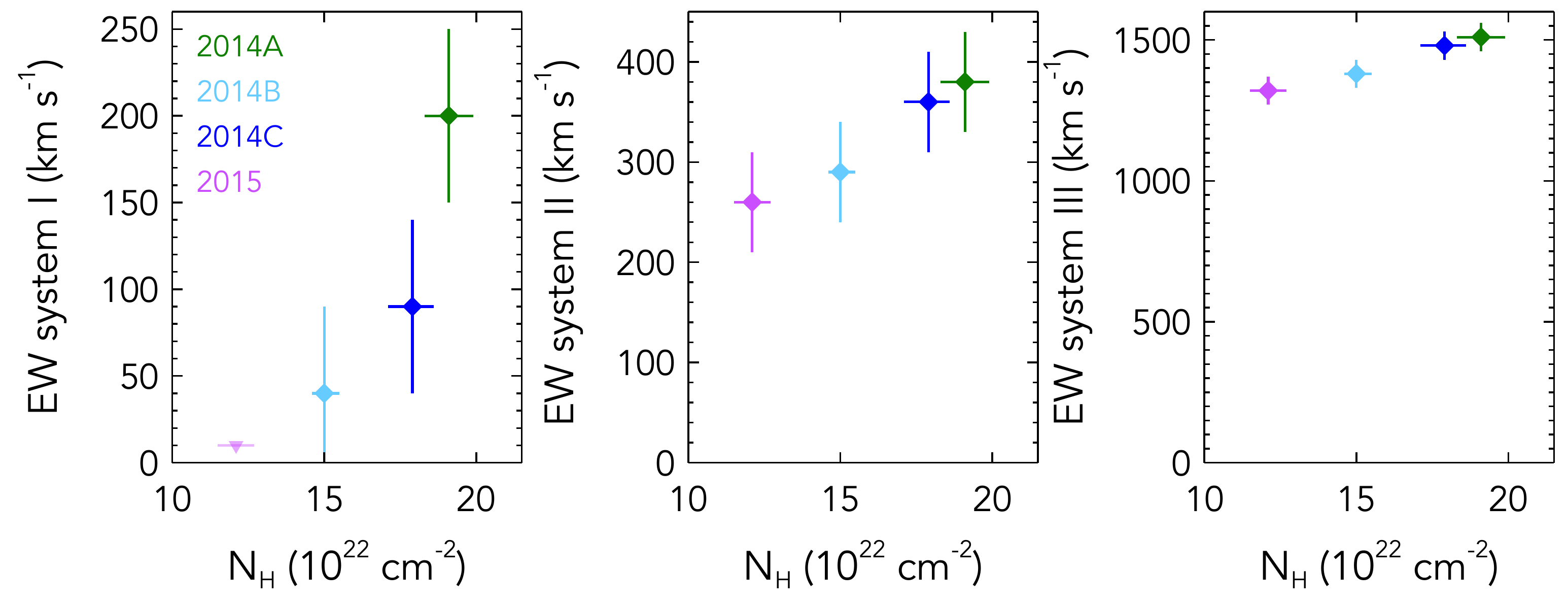}
   \caption{EW of the {\CIV} absorption systems against the column density of the X-ray partially covering absorber in each epoch of coordinated \textit{XMM-Newton/HST} observation of PG 1126-041. From left to right: system I, system II, and  system III (system I was not detected during 2015). }
   \label{FIGEWNH}%
    \end{figure*}

In the X-ray band, we detected two absorbers: one highly ionized ultra-fast outflow absorbing mainly in the Fe K band ($E\sim 7-10$ keV); and one partially covering the X-ray continuum emission source and at a lower ionization state, with the largest opacity between $E\sim 0.5-2$ keV. While both absorbers are found to be variable with time, the spectral variability observed in PG 1126-041 is dominated by variations in the column density of the partially covering absorber (Sect. \ref{sec:partcov}).
In the UV band, we focused on the absorption in the {\CIV} region, which shows three systems of blueshifted absorption lines with maximum velocities between $-3300$ and $-6700$ km s$^{-1}$.
These systems show variability in a coordinated way between different epochs, being strongest during 2014A and 2014C and weakest during 2014B and 2015.

The variability observed in the {\CIV}-absorbing wind appears to be coordinated with the column density variations of the  X-ray partially covering absorber.
The 2014A and 2014C epochs are the ones with the largest X-ray-absorbing column density of the partially covering gas ($N_H\sim 1.8-1.9\times 10^{23}$ cm$^{-2}$).
The X-ray partial covering absorber column density is a factor of 20\% and 40\% smaller in 2014B and 2015, respectively, than in 2014A and 2014C. Concurrently, the {\CIV} absorption line profiles are much more similar and weaker in 2014B and 2015 compared to 2014A and 2014C.
We plot in Fig.~\ref{FIGEWNH} the EW of each {\CIV} absorption system against the median of the posterior probability of the X-ray partially covering absorber column density for each epoch of observation, computed with the \texttt{[(partcov*xstar500)*(fast32*pow)]} model.
This coordinated variability strongly suggests that the X-ray partially covering absorber and the   highly blueshifted {\CIV} absorbers are connected.
The UV absorption troughs observed in BAL QSOs are often inferred to be shaped by partially covering absorption of the UV continuum source, and the covering fractions are comparable to those obtained for the X-ray partially covering absorber in PG 1126-041 \citep[$C_f \sim 80-90\%$, see e.g.,][]{1999ApJ...516...27A, 2011MNRAS.411..247R}.
This might suggest similar geometries for the UV-absorbing gas and the X-ray-absorbing gas relative to their sources of continuum emission.

It would be interesting to compare the velocities of the outflowing absorbers. Given the limited spectral resolution of the EPIC detectors, the multitude of closely-spaced spectral transitions produced in the $\log\xi\sim 2$ partially covering absorber are not individually resolved in our X-ray observations, and a velocity shift cannot be measured.
On the contrary, the UFO with $\log\xi\sim 3.5$ only imprints a few strong transitions in the Fe K band, which can then be used as a velocity shift marker even at moderate spectral resolution. The velocity projected along the line of sight of the UFO has a blueshift of $\sim 0.06c$ (Table~\ref{table:fitALLGAUSS}), a factor of three larger than the maximum outflow velocity projected  along the line of sight for the UV absorber. When the UFO is modeled within the accretion disk wind scenario, the biconical
geometry assumed for the wind gives an even larger deprojected velocity of $\sim -0.2 c$. In order to compare this value with the UV maximum velocity, a geometry for the UV absorbing wind should also be assumed.

One possibility is that, from farther out to closer in toward the central SMBH of PG 1126-041, our line of sight goes through the UV emitting and absorbing region, then encounters the X-ray partially covering absorber, then the X-ray UFO, and finally the intrinsic continuum source.
We might therefore be observing an innermost wind, where the highly ionized ($\log\xi\sim 3.5$) X-ray absorption is produced and the matter is accelerated (up to $\upsilon_{\infty}\sim -0.2c$), together with the portion of the outflow situated further out of the acceleration zone, which is thermally unstable and therefore clumpy and produces the X-ray partially covering absorption \citep{2022ApJ...931..134W}.
The X-ray partially covering absorber then acts as a 'filter' (a variable patchy screen) of the incoming ionizing soft X-ray photons for the UV-absorbing gas \citep{2007ApJ...660..152M}.
The UV absorber is exposed to a number of X-ray photons proportional to $ (1-C_f) + C_f\, e^{-\tau} $, where the first term is the unobscured photon flux and the second term is the photon flux making it through any shielding gas.
Generally speaking, the closer to the SMBH is the wind launching point, the fastest must be its terminal velocity. Assuming that the observed velocities are proportional to the wind terminal velocity, the following scenario might explain the observations:
when more massive clumps along the line of sight are covering the X-ray continuum source (e.g., during 2014A and 2014C), a lower number of X-ray photons are reaching the {UV}-absorbing wind.
When the X-ray absorption along the line of sight is reduced (e.g., during 2014B and 2015), then the  X-ray flux reaching the {\CIV}-producing region is larger, thus overionizing the UV-absorbing wind.

Hints of the nature of the X-ray partially covering absorber come from the existence of an anti-correlation between its column density and covering fraction in different epochs, as shown in the top panel of Fig.~\ref{FIG9}.
One possibility is that the $1-10\%$ fraction of X-ray light that escapes unaffected by the partially covering absorber is in fact scattered light. As the column density increases, so does the electron scattering optical depth, which allows more photons from the background (continuum) source to scatter off the electrons of the absorber without changing their spectrum. Scattering might take place within the  partially covering absorber itself and in the accretion disk wind.
One way to discriminate between these two scenarios would be using X-ray polarimetric observations: in fact, the flux between $\sim 0.6-1$~keV is expected to be polarized if it is produced by electron scattering, while this is not the case for directly transmitted flux.

Similarly to PG 1126-041, the blueshifted absorption line system in the UV spectrum of the AGN HS 1603+3820 was shown to display coordinated variability on time scales of years \citep{2007ApJ...660..152M}, which could be explained either by electron scattering or by the effects of a variable clumpy `screen' between the mini-BAL gas and the continuum source. The latter hypothesis was supported by UV spectropolarimetric observations \citep{2010ApJ...719.1890M}.
The physical characteristics of the X-ray absorbers in sources with intrinsic blueshifted UV absorption are not clear. For example, \citet{2013MNRAS.435..133H} conducted a study of eight AGN with high-velocity (outflow velocities $\sim 0.1-0.2c)$ UV-absorbing winds, finding no significant X-ray absorption. They suggest that the UV absorber ionization is low, not due to shielding gas but due to high densities likely resulting from magnetic confinement. However, only cold (neutral) X-ray absorption was considered, and the limits in neutral column density found ($N_H\sim 0.3-5\times 10^{22}$ cm$^{-2}$) might  be consistent with larger column densities of gas along the line of sight for more complicated scenarios such as ionized or partially covering X-ray absorption. On the basis of the observed spectral complexities, it has been shown by \citet{2016AN....337..459G} that the X-ray spectra of AGN with BAL and mini-BAL features are consistent with large column densities $N_H=10^{22-24}$ cm$^{-2}$ of gas along the line of sight, if the ionization state is significantly different from neutral, or if the gas is only partially covering the X-ray continuum source.

\subsection{PG 1126-041 in the grand scheme of AGN\label{discu:pg}}

PG 1126-041 is a low-redshift AGN with powerful nuclear winds and is thus similar to BAL QSOs, which usually have a much lower X-ray flux and therefore a lower number of detected photons at the detectors, which prevents complex spectral models to be applied.
The general observed spectral properties of PG 1126-041 and similar sources might therefore be used to infer the general properties of large samples of sources with a much lower number of X-ray photons available, thus paving the way for future studies with larger X-ray telescopes.

The partial obscuration of the central X-ray continuum emission source by large column densities of ionized gas is common in AGN, and in fact, partially covering absorbing gas might be an AGN ingredient on all scales.
Partially covering X-ray absorption by cold, Compton-thick clouds transiting on torus or BLR-scales ($\sim 10^{3}-10^6\,r_g$) has been detected in several AGN, notably  NGC 1365 \citep[][]{2005ApJ...623L..93R,2009MNRAS.393L...1R,2010A&A...517A..47M,2015ApJ...804..107R}, Mrk 766 \citep{2011MNRAS.410.1027R}, H0557-385 \citep{2009MNRAS.394L...1L}. Thanks to long temporal baseline observations of a large sample of AGN, the occurrence of such occultation events has been discovered to be relatively common \citep{2014MNRAS.439.1403M}.
Following deep monitoring studies, several historically unabsorbed AGN have been recently discovered that are affected by transient obscuration events, which depress the observed X-ray  flux and where blueshifted UV absorption lines are observed to emerge \citep[e.g.,][]{2014Sci...345...64K,2016A&A...586A..72E,2017A&A...607A..28M,2019A&A...621A..12K,2021A&A...652A.150M}. Partially covering X-ray absorbing gas has also been detected very close to the central SMBH, as inferred by the rapid variability, in the luminous AGN PDS 456 \citep[at tens to hundreds $r_g$, e.g.,][]{2015Sci...347..860N,2016MNRAS.458.1311M,2018ApJ...867...38R}.

In the case of PG 1126-041, we can use the time variability  constraints between the closely spaced 2014 observations to get an estimate of the distance of the partially covering X-ray absorber from the continuum source.
The velocity of the partially covering absorber cannot be constrained with our data, but we can consider the range in velocities given by the most blueshifted UV absorber ($-0.02c$) and the deprojected X-ray UFO ($-0.2c$).
If one assumes that the transverse velocity is comparable to the radial velocity, so that $\Delta R$ is the distance the absorber moves during time $\Delta t$, then the variations observed on time scales of about 10 days would correspond to a thickness $\Delta R\sim 5\times 10^{14-15}$ cm. Given $n\sim N_H/\Delta R$, for $\langle N_H\rangle \sim 1.3\times 10^{23}$ cm$^{-2}$ one would obtain $n \sim 2.6\times 10^{7-8}$ cm$^{-3}$. Inserting these density estimates into the ionization parameter definition $\xi=L/nR^2$ and using $L=10^{44}$ erg s$^{-1}$ and $\log\xi=2$, one would obtain a distance for the partially covering absorber $R_{PC}\sim (6.2-20)\times 10^{16}$ cm. This distance corresponds to about $3500-11000\, r_g$ for a black hole mass estimate of $M_{BH}=1.2\times 10^8 M_{\odot}$ \citep{2007ApJ...657..102D}.

Assuming that the variations in column density are due to a single cloud moving across the line of sight, another estimate of the location of the X-ray partially covering absorber can be placed by calculating the absorber transverse velocity necessary to move across the X-ray emitting region during the shortest separation between consecutive observations and comparing it to the orbital (Keplerian) distance. The size of the X-ray emitting region in PG 1126-041 is constrained to be $<13\,r_g$ by \citet{2011A&A...536A..49G}, using the duration of a continuum flare of 8 ks. An absorber with a transverse velocity of $\upsilon_K\sim 0.008\,c$ would cross such a region during the 11 days elapsed between 2014A and 2014B, and this velocity would correspond to orbits at $\sim 15,000\,r_g$ from the central SMBH. An absorber with a higher velocity would of course cross the same region in a shorter time, e.g., in about 4.4 days at the velocity of the UV absorber and in about 10.5 hours at the velocity of the X-ray UFO.
The partially covering absorber would therefore be on the broad line region scales, consistent with the observed variations in UV-absorbing gas coordinated with variations in X-ray absorbing column density.

For comparison, the location of the UFO can be estimated in an analogous way using the variations observed on time scales of about 8 ks during 2009 \citep{2011A&A...536A..49G}, an average column density $5\times 10^{23}$ cm$^{-2}$ and column density variations $\Delta N_H\sim 2.5\times 10^{23}$ cm$^{-2}$, obtaining a distance estimate $R_{UFO}\sim 30-40\,r_g$.
The distance of the UFO to the X-ray continuum source might therefore be comparable to the size of the latter; this would have implications for the detailed theoretical modeling of the X-ray UFO, as in this case the important assumption of a point-source continuum as seen by the absorber could not be valid anymore.

In any case, our distance estimates should be considered simple approximations, as in our calculations we neglected the presence of the wind itself, with all the physical consequences this might have: notably a different (than spherical) geometry and a highly dynamical environment, where the effects of pressure and temperature gradients across the flow are expected to generate much more complex observational scenarios \citep[e.g.,][]{2012ApJ...758...70G, 2021ApJ...914..114G,2022ApJ...931..134W}.
What can be said with our distance estimates is that the X-ray partially covering absorber either lies between the UFO and the UV-absorbing gas, or it is co-spatial (within 1.5 light hours, the minimum time separation between \textit{XMM-Newton} and HST-COS observations of PG 1126-041) with the latter.
Dynamical thermal instability has been recently demonstrated to be a mechanism that can make AGN winds clumpy beyond the acceleration zone \citep{2022ApJ...931..134W}.

The X-ray obscurer detected in NGC 5548 has a comparable covering fraction to the X-ray partial covering absorber in PG 1126-041, but a significantly lower column density and ionization state \citep[$N_H\sim 1.2\times 10^{22}$ cm$^{-2}$ and $\log\xi\sim -1.2$,][]{2014Sci...345...64K}.
The column density and the ionization state of the partially covering absorber in PG 1126-041 are more similar to those of the obscurer detected in NGC 3783 \citep{2017A&A...607A..28M,2020A&A...634A..65D,2022A&A...659A.161C}, although the covering fraction in the latter AGN is much lower (by about a half) than the covering fraction in PG 1126-041.
A good match in covering fraction and column density is found between the X-ray partial covering absorber in PG 1126-041 and the X-ray obscurer in Mrk 817 \citep{2021ApJ...922..151K}, which shows variability of column density on time scales of days and weeks and coordinated intensity of the phosphorus ion P~\textsc{v} absorption troughs similar to the one observed in the {\CIV} absorption troughs in PG 1126-041.
Therefore, the X-ray partially covering absorber in PG 1126-041 has the characteristics of a massive, highly ionized X-ray obscurer.

The general X-ray spectral properties of PG 1126-041 are also remarkably similar to those of PDS 456, the most well-studied luminous QSO hosting an X-ray-absorbing accretion disk wind \citep{2003ApJ...593L..65R}.
PDS 456 is absorbed by variable ionized gas and has one or more UFOs with observed velocities $\upsilon_{out}=-(0.25-0.3)c$ \citep{2009ApJ...701..493R,2018MNRAS.476..943H}.
The observed velocity shift in PG 1126-041 is much smaller, $\langle\upsilon_{out}\rangle \sim - 0.06c$, however $\upsilon_{out}$ is the projected velocity along the line of sight, therefore a lower limit on the actual wind velocity. When taking into account the geometry of the accretion disk wind, a terminal velocity $\langle\upsilon_{\infty}\rangle \sim -0.2c$ was recovered for the accretion disk wind of PG 1126-041.
Therefore, the differences in observed velocity shift in PDS 456 and PG 1126-041 might be mainly due to a different inclination angle of our line of sight with respect to the wind: in PDS 456 the inclination angle is smaller, of the order of $50^{\circ}$ \citep{2022MNRAS.515.6172M}, compared to the $80^{\circ}$ derived for PG 1126-041.
Recently, an even larger component with $\upsilon_{out}=-0.45c$ has been unveiled in the X-ray spectrum of PDS 456 using a broadband (joint \textit{XMM-Newton} + \textit{NuSTAR}) observation \citep{2018ApJ...854L...8R}.
Future hard X-ray observations of PG 1126-041 should help clarify whether there is a higher-velocity UFO phase in this AGN as well.
In any case, it is likely that the accretion disk wind in PG 1126-041 is a scaled-down version of the extremely powerful wind of PDS 456 in terms of energetics.
The black hole mass estimate for PDS 456 is about one order of magnitude larger than for PG 1126-041, therefore all the time scales of variability in PG 1126-041 should be more rapid by a factor of $\sim 10$. Given the observed variability of the X-ray partially covering absorber in PDS 456 on time scales of 100 ks \citep{2016MNRAS.458.1311M}, it could be worth investigating whether such a component varies on time scales of $\sim 10$ ks in PG 1126-041.

\section{Conclusions\label{sec:CONCLU}}

The results of the analysis of eight \textit{XMM-Newton} observations of PG 1126-041 between 2004 and 2015, the last four of which were taken quasi-simultaneously with \textit{HST}-COS exposures, are the following:
\begin{description}
\item[$\bullet$] The spectral shape of PG 1126-041 between $0.3-10$ keV is complex, showing strong reprocessing of the intrinsic X-ray emission:  a fit to a phenomenological power law gives a very flat photon index $\Gamma\sim 0.6$; however, once accounting for complex absorption along the line of sight, a more typical $\Gamma\sim 1.9$ is measured.\\
\item[$\bullet$] A massive X-ray partially covering absorber is detected in all epochs, independent of the underlying broad-band continuum modeling.
The ionization parameter is $\log\xi\sim 2$, the covering fraction $C_f\sim 95\%$, and the column density is in the range $N_H\sim (5-20)\times 10^{22}$~cm$^{-2}$.\\
\item[$\bullet$] X-ray spectral variability of PG 1126-041 is observed between every epoch probed, with time separations as short as 11 days; the spectral variability is driven by the variable column density of the partially covering absorber.\\
\item[$\bullet$] The column density of the X-ray partially covering absorber shows coordinated variability with the highest-velocity components of the {\CIV} absorber observed in the UV with the COS during the last four epochs of observation.
In particular, during 2014A and 2014C the X-ray absorbing column density was maximum and the {\CIV} absorber had the maximum equivalent width; during 2014B and 2015 the column density was smaller by respectively 20\% and 40\%, and the {\CIV} absorber had the minimum equivalent width. \\
\item[$\bullet$] In addition to the partially covering absorber, a highly ionized outflowing absorber (UFO, detected with a statistical confidence $>99.9\%$ in 3/8 epochs) and a broad emission feature modeled with a phenomenological Gaussian emission line centered at $E=5.45$ keV and with a $\sigma\sim 1$ keV (present in half of the epochs) are needed to reproduce the spectral shape of PG 1126-041 in the Fe K band. Both components can be accounted for by a biconical accretion disk wind model observed along an equatorial line of sight.\\
\item[$\bullet$] When the UFO is reproduced by 1D photoionization models, a column density $\sim 6\times 10^{23}$ cm$^{-2}$, an ionization parameter $\log\xi\sim 3.5$, and a velocity projected along the line of sight $\upsilon_{out}\sim -0.06\,c$ are measured. When the UFO is modeled within the accretion disk wind scenario, the biconical geometry assumed for the wind gives a deprojected terminal velocity $\upsilon_{\infty}\sim -0.2\,c$.\\
\item[$\bullet$] While the accretion disk wind model used ignores the effects of gas pressure in computing the physical structure of the outflow and assumes a very simple geometry of a thin cone, this is a necessary step toward a more realistic treatment of AGN accretion disk winds.
For example, it takes into account both scattering and absorption of X-ray photons, and the measured intrinsic X-ray  luminosity is therefore larger (by a factor of $\sim 4\times$) compared to the one measured using 1D photoionization codes, which only consider the effects of absorption of photons along the line of sight.\\
\item[$\bullet$] Assuming a launching radius for the wind inversely proportional to the observed velocity, a possible scenario to interpret the XMM-Newton/HST-COS observations of PG 1126-041 is the following: a wind with an inner ultrafast component launched at a few tens $r_g$ from the central SMBH (the UFO) fragments beyond its acceleration zone due to thermal instability;
the subsequent clumps that form in the flow might be the X-ray partially covering absorber, which might be co-located (or at distances shorter than 10 light-days) with the UV-absorbing wind. The X-ray partially covering absorber clumps act as a variable screen between the X-ray photons transmitted through and scattered off the wind and the UV absorber. Therefore, in epochs of low  X-ray column density the highest velocity, innermost UV-absorbing wind would get destroyed by the large flux of X-ray photons able to reach it.\\
\item[$\bullet$] Accretion disk wind models will need to account for the presence of massive clumps in their structure at some point in the future.
\end{description}

This work adds PG 1126-041 to the growing number of AGN studied with parametrized accretion disk wind models based on radiative transfer calculations. These studies can allow us to estimate physical parameters of the inner accretion and ejection flow around SMBHs and are a necessary step toward the development of full-hydrodynamical disk wind models that might explain all the observed complexities.
Coordinated multiwavelength spectroscopic analyses of AGN are very promising in constraining the dynamical properties of their accretion disk winds.
The line of sight toward PG 1126-041 offers a privileged view through a highly dynamical accretion disk wind, and it is worth exploring with future studies in order to shed light on the general connection between the accretion and the ejection flow around SMBHs.

\begin{acknowledgements}
We thank the referee for the comprehensive and constructive review that helped improve the presentation of the results.
 MG is supported by the ``Programa de Atracci\'on de Talento'' of the Comunidad de Madrid, grant number 2018-T1/TIC-11733.
P.R.H. acknowledges support from start-up funding provided by UW Bothell, from start-up funding from the College of Natural
Resources and Sciences, and from the Sponsored Programs Foundation at Humboldt State University (now Cal Poly Humboldt) through a Research, Scholarly, \& Creative Activity grant. GP acknowledges funding from the European Research Council (ERC) under the European Union’s Horizon 2020 research and innovation programme (grant agreement No 865637).
JNR acknowledges support from NASA cooperative grant 80NSSC22K0474.
 VB acknowledges support from NASA cooperative grant 80NSSC22K0220.
 We thank Stuart S. Sim for making his disk wind model available to help us interpret our observations, Jeremy Sanders for writing \texttt{Veusz}, Craig Gordon and Keith Arnaud for developing \texttt{pyXspec}, and Johannes Buchner for developing \texttt{UltraNest} and \texttt{BXA} and for kind and helpful discussions.
 Based on observations obtained with \textit{XMM-Newton}, an ESA science mission with instruments and contributions directly funded by ESA Member States and NASA. We acknowledge financial support from NASA/\textit{XMM-Newton} via the grant NNX10AE11G.
This research is based on observations made with the NASA/ESA \textit{Hubble Space Telescope} obtained from the Space Telescope Science Institute, which is operated by the Association of Universities for Research in Astronomy, Inc., under NASA contract NAS 5–26555. These observations are associated with programs 13429 and 13836.
 This research has made use of NASA's Astrophysics Data System Bibliographic Services.

 \end{acknowledgements}


\bibliographystyle{aa}
\bibliography{biblio}

\clearpage
\begin{appendix}
\section{Bayesian analysis results}
\begin{figure*}[h!t]
  \centering
\includegraphics[width=18cm]{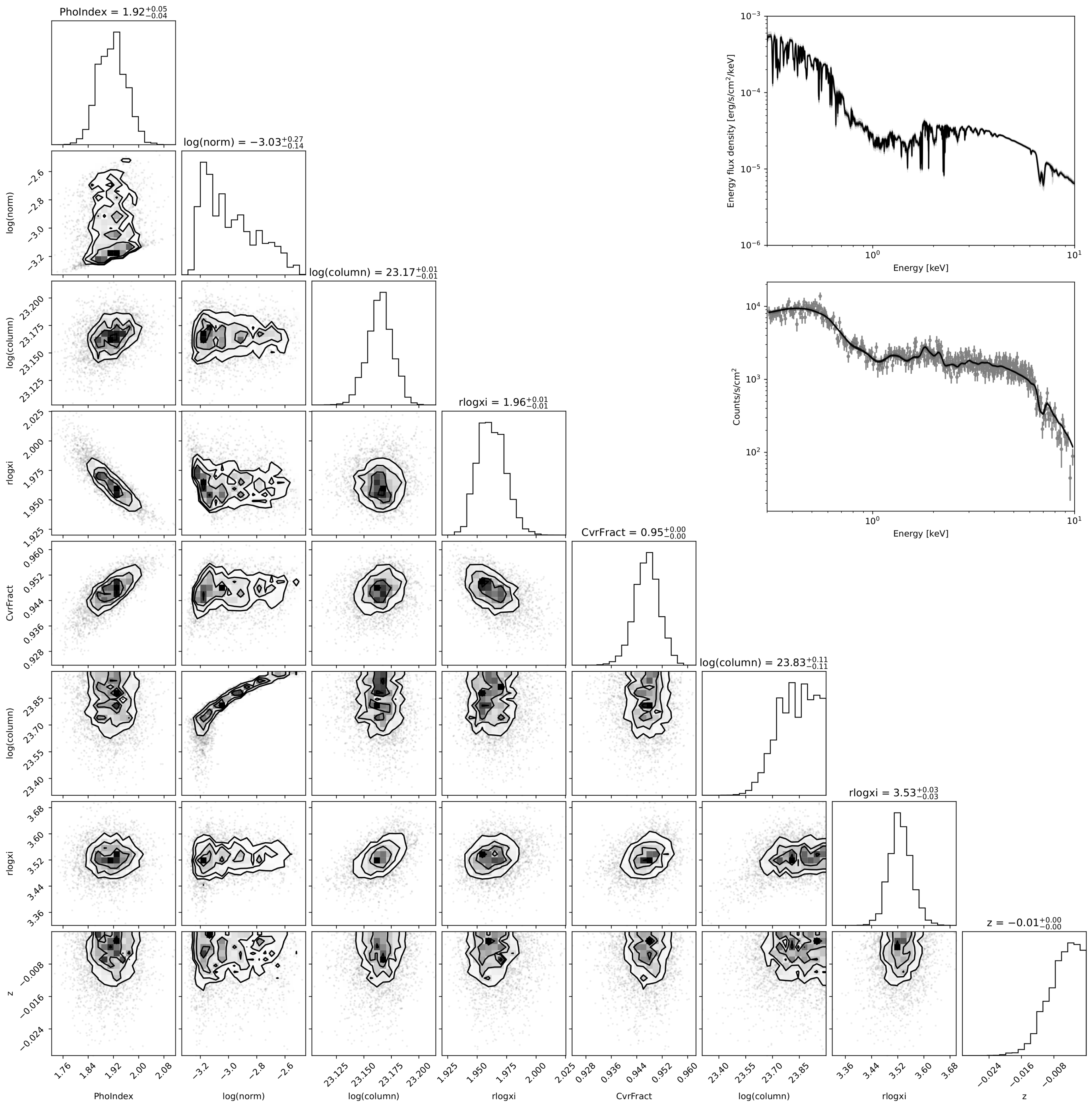}
   \caption{Bayesian analysis results for the parameters of the model \texttt{[(partcov*xstar500)*(xstar5000*pow)]} given the 2009 data. The corner plot reports the one-dimensional (histograms) and two-dimensional (credible contours) posterior probability distributions for the model parameters. The top right panels report the posterior probability distributions of the theoretical model (top) and of the model convolved with the instrumental response, with the binned EPIC-pn data overplotted (bottom).}
   \label{FIGbxa2009xstar}%
    \end{figure*}

\begin{figure*}[hb]
  \centering
\includegraphics[width=18cm]{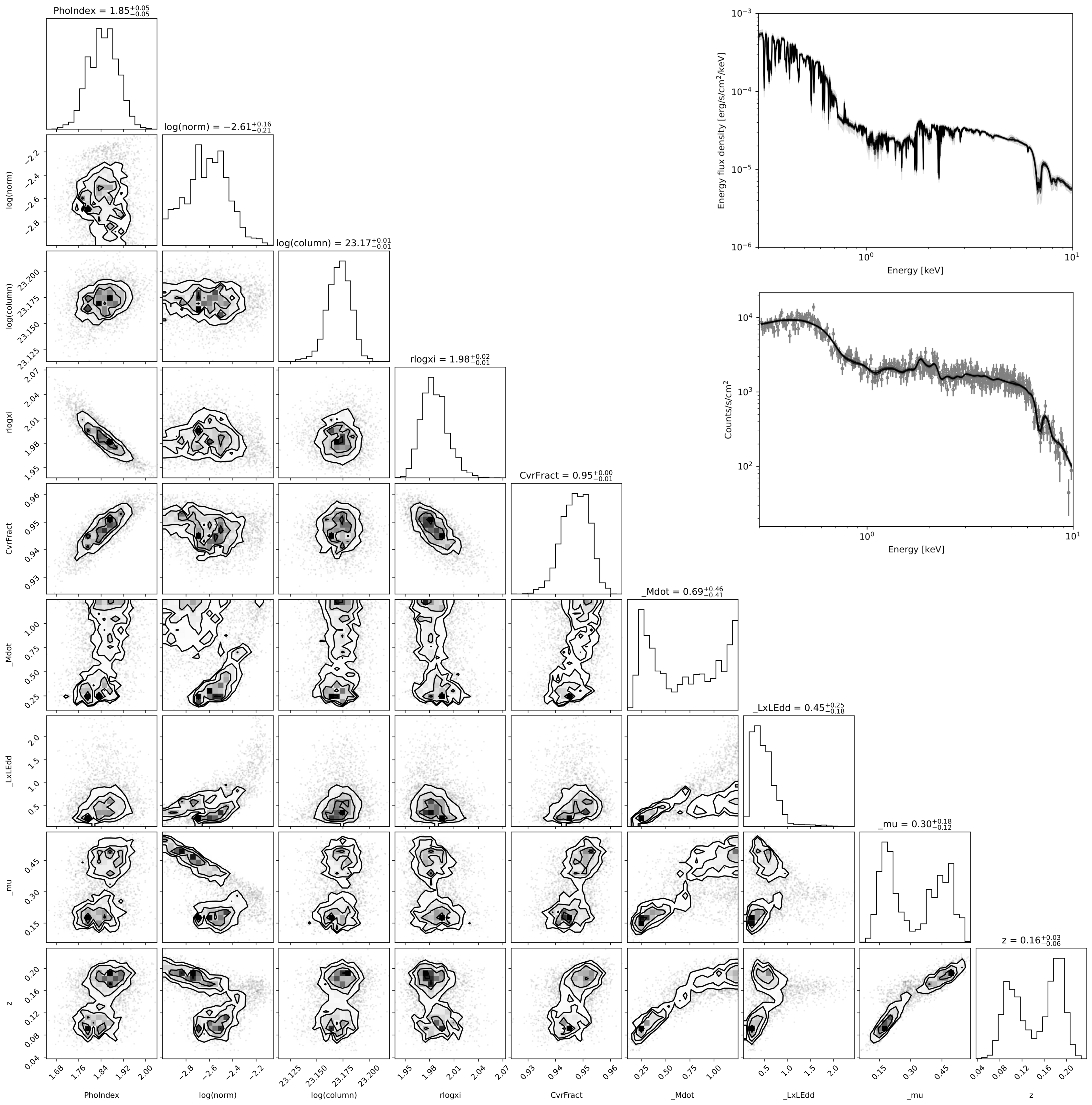}
   \caption{Bayesian analysis results for the parameters of the model \texttt{[(partcov*xstar500)*(fast32*pow)]} given the 2009 data. The corner plot reports the one-dimensional (histograms) and two-dimensional (credible contours) posterior probability distribution for the model parameters. The top right panels report the posterior probability distributions of the theoretical model (top) and of the model convolved with the instrumental response, with the binned EPIC-pn data overplotted (bottom).
   }
   \label{FIGbxa2009}%
    \end{figure*}

We report an example of the results of the Bayesian analysis described in Sect.~\ref{sec:partcov} referred to the 2009 data. In Fig.~\ref{FIGbxa2009xstar}, we report the posterior probability distributions for the model \texttt{[(partcov*xstar500)*(xstar5000*pow)]}. In Fig.~\ref{FIGbxa2009}, we report the posterior probability distributions for the model \texttt{[(partcov*xstar500)*(fast32*pow)]}.

\section{Normalization of the HST-COS spectra\label{AppendixB}}

\begin{figure*}[hb]
 \centering
 \includegraphics[width=9cm]{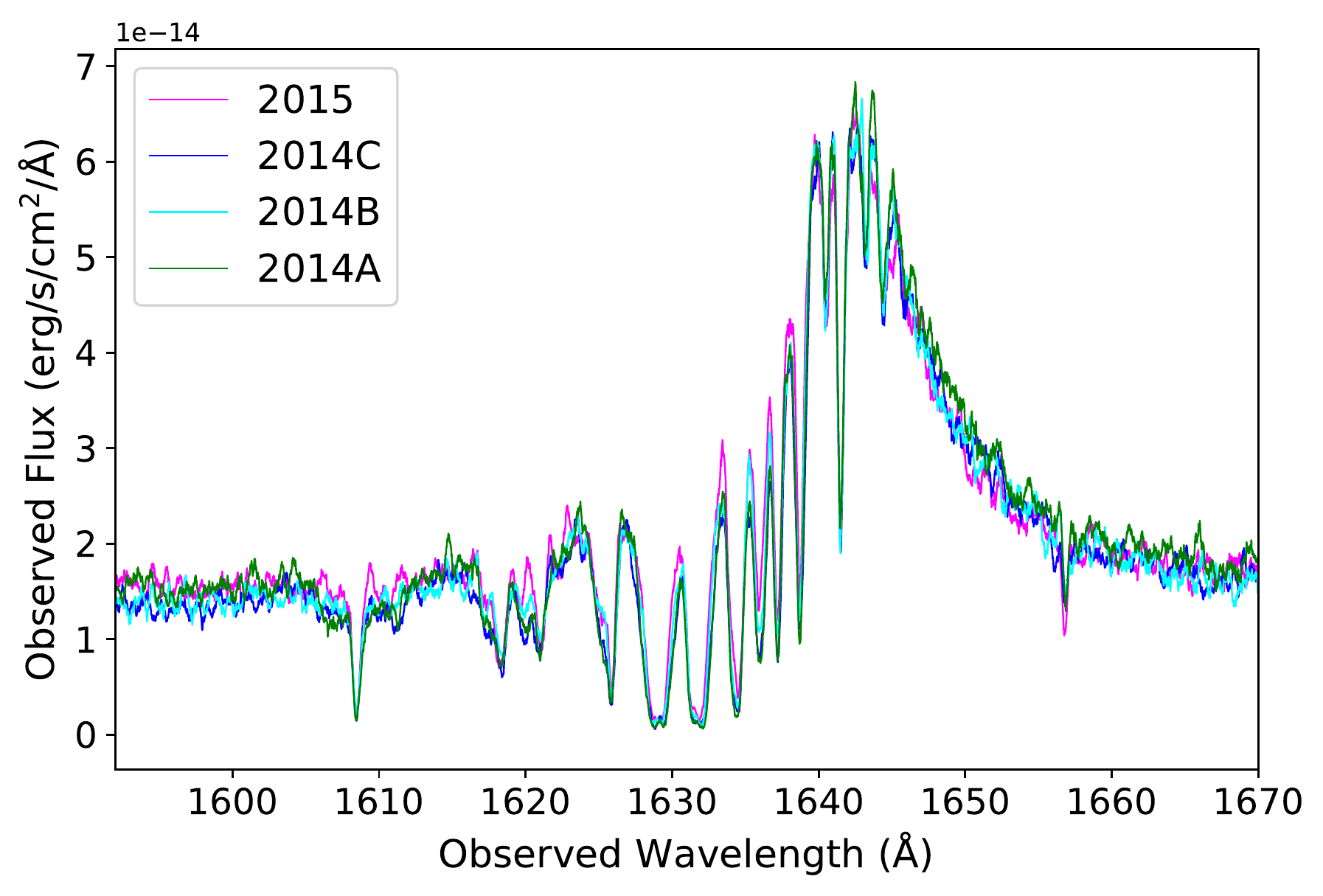}
 \includegraphics[width=9cm]{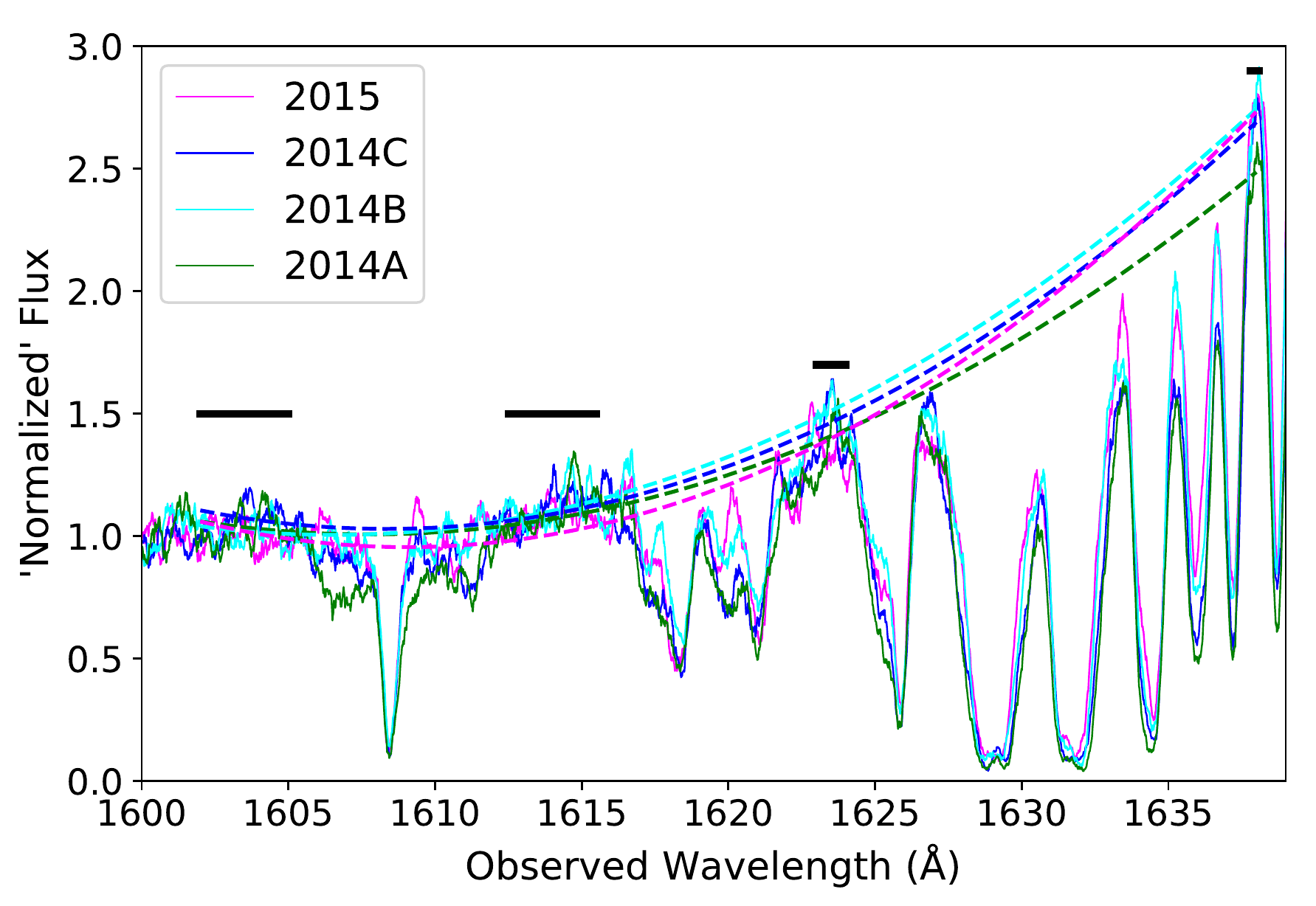}
   \caption{Normalization procedure for the {\CIV} region of the \textit{HST}-COS spectra of PG 1126-041 taken between 2014 and 2015. Left panel: observed spectra before normalization. Right panel:
 second-order polynomial fits to the region of interest where the absorption is present. The black horizontal lines show the regions selected to anchor the fit.}
   \label{FIGUVMODEL}%
    \end{figure*}

We show the normalization procedure for the four \textit{HST}-COS epochs of observation of PG 1126-041 described in Sect.~\ref{sec:XUV}.
The left panel of Fig.~\ref{FIGUVMODEL} shows the four \textit{HST}-COS original spectra before the normalization process. All of them are very similar in their continuum. The right panel of Fig.~\ref{FIGUVMODEL} shows the second-order polynomial fits to the region of interest where the absorption is present. The black horizontal lines show the regions selected to anchor the fit, that were the same for all spectra: $1602-1605\,\AA$, $1612.5-1615.5\,\AA$, $1623-1624\,\AA$, and $1637.8-1638.1\,\AA$. All the spectra in both figures have been smoothed with a boxcar filter with a width of 31 pixels (corresponding to less than 0.4 \AA\, or 75 {\kms}) for visualization purposes.

 \end{appendix}
\end{document}